\documentclass[aps,reprint,groupedaddress,showpacs]{revtex4-1}
\usepackage{amsmath,bm}
\usepackage{amssymb}
\usepackage{color}
\usepackage{multirow}
\usepackage[colorlinks,bookmarks=false,citecolor=red,linkcolor=blue,urlcolor=blue]{hyperref}
\usepackage{graphicx}
\usepackage{mathrsfs,bbm}
\usepackage[caption=false]{subfig} 
\usepackage{wrapfig}
\usepackage[version=3]{mhchem}

\def\a{{\bf a}}
\def\aidag{{a_{i}^{\dagger}}}
\def\bidag{{b_{i}^{\dagger}}}

\def\k{{\bf k}}
\def\r{{\bf{r}}}
\def\shat{\hat{s}}
\def\that{\hat{t}}
\def\sbar{\bar{s}}
\def\Qhat{\hat{Q}}
\def\Phat{\hat{P}}
\def\onebar{{\bar{1}}}
\def\q{{\bf q}}
\def\MC{\mathcal{C}}
\newcommand*\afdn{\includegraphics[width=5mm]{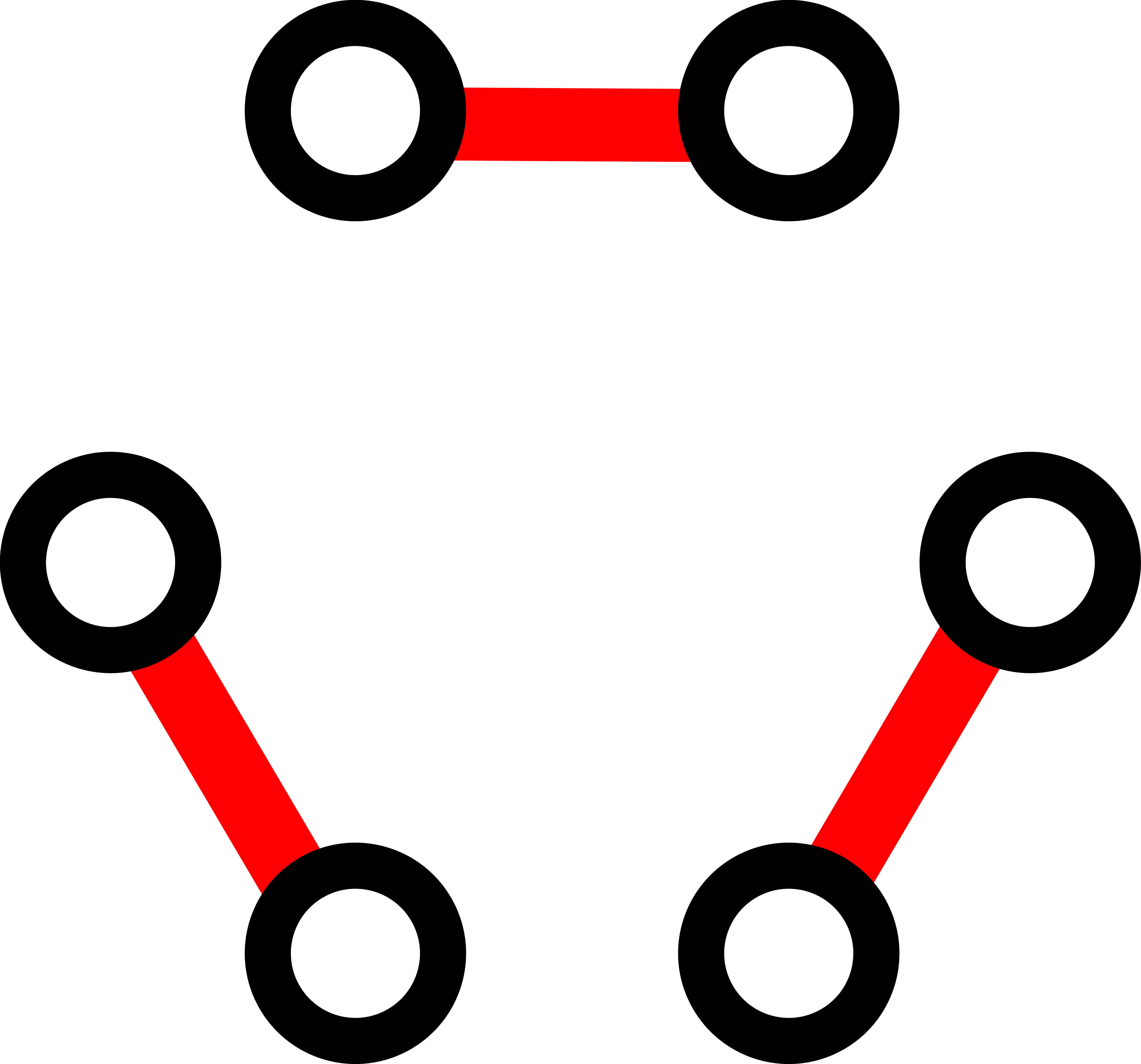}}
\newcommand*\afup{\includegraphics[width=5mm]{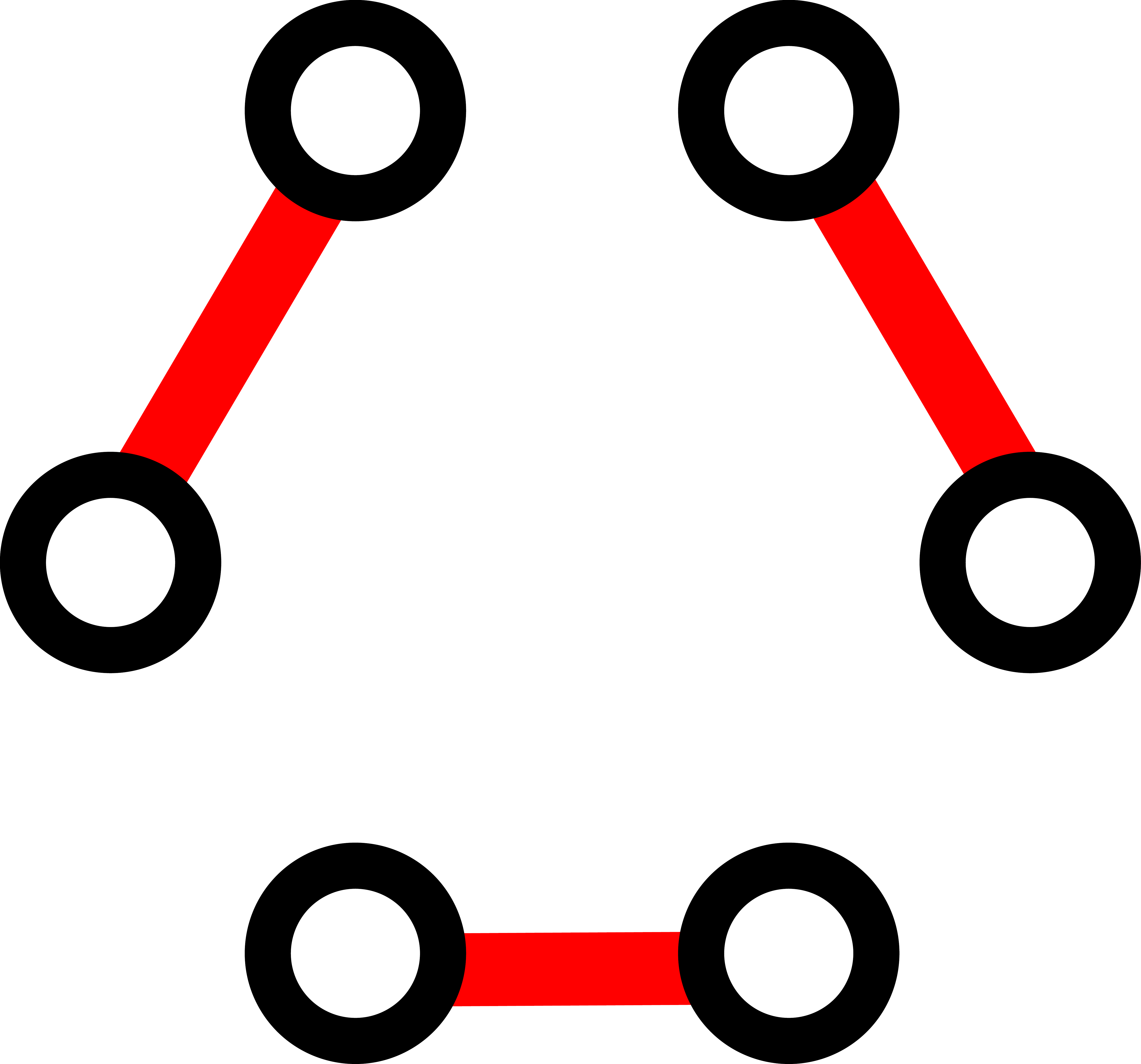}}
\newcommand*\fer{\includegraphics[width=5mm]{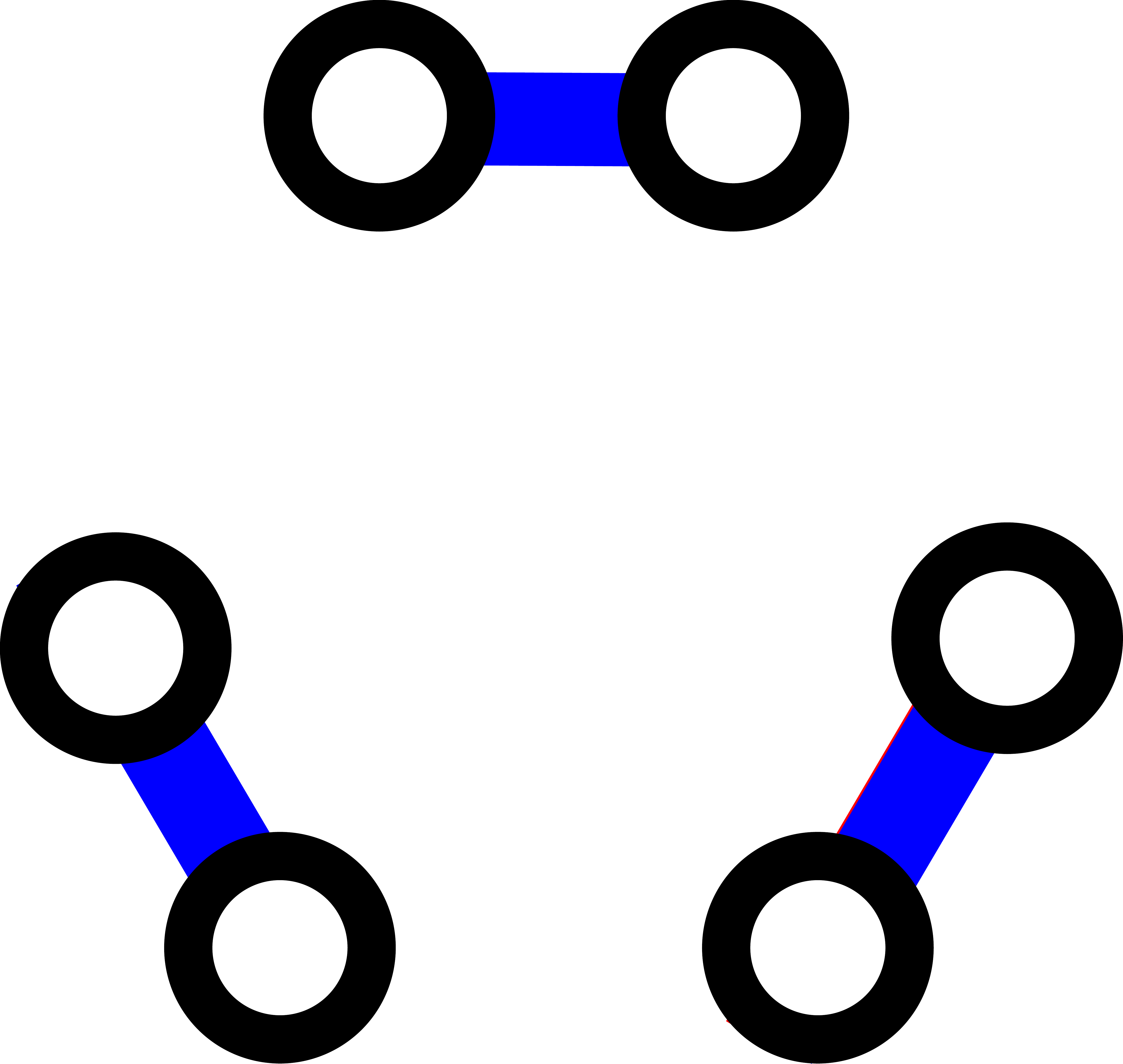}}
\newcommand*\afer{\includegraphics[width=5mm]{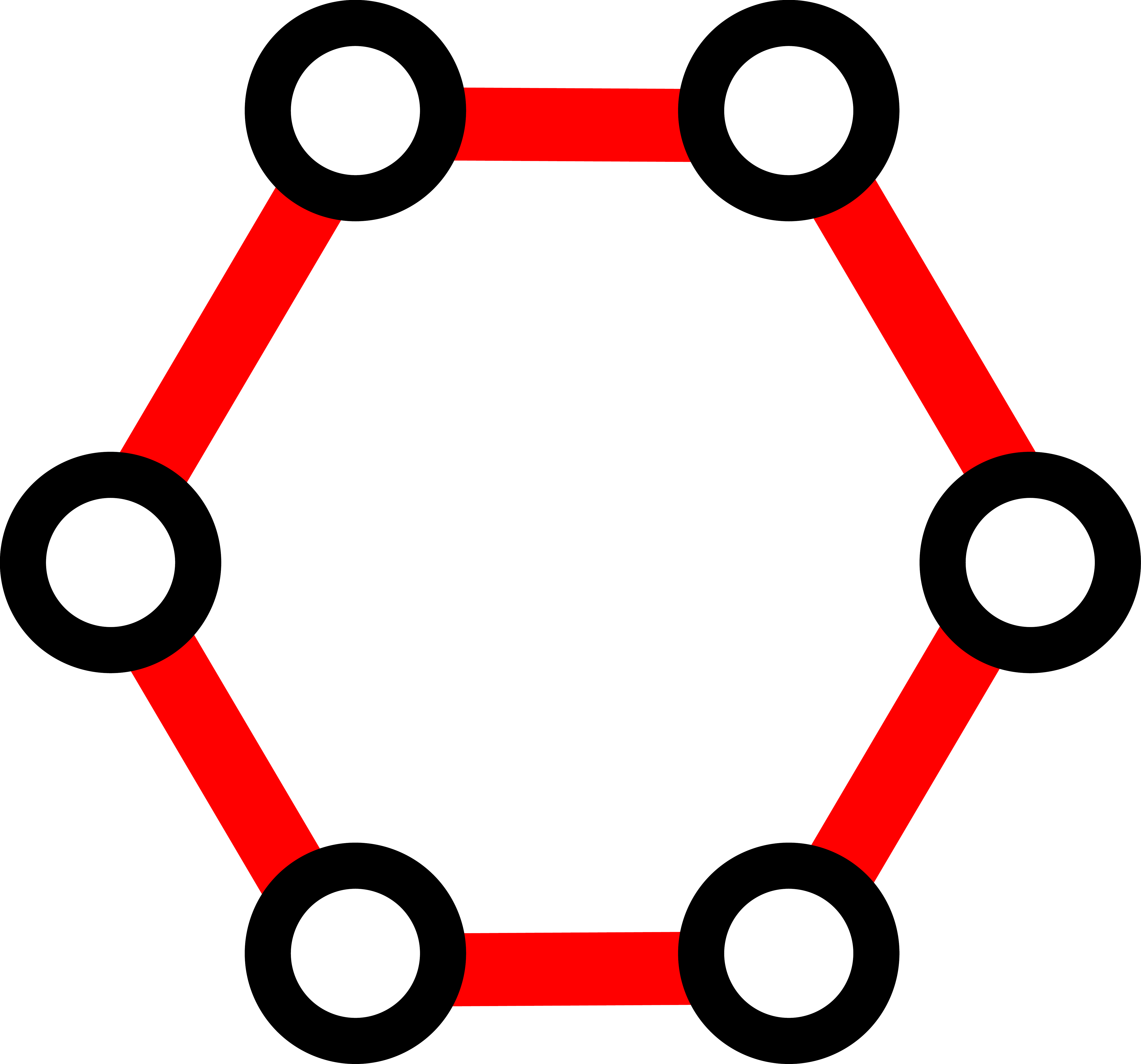}}
\newcommand*\afdhss{\includegraphics[width=5mm]{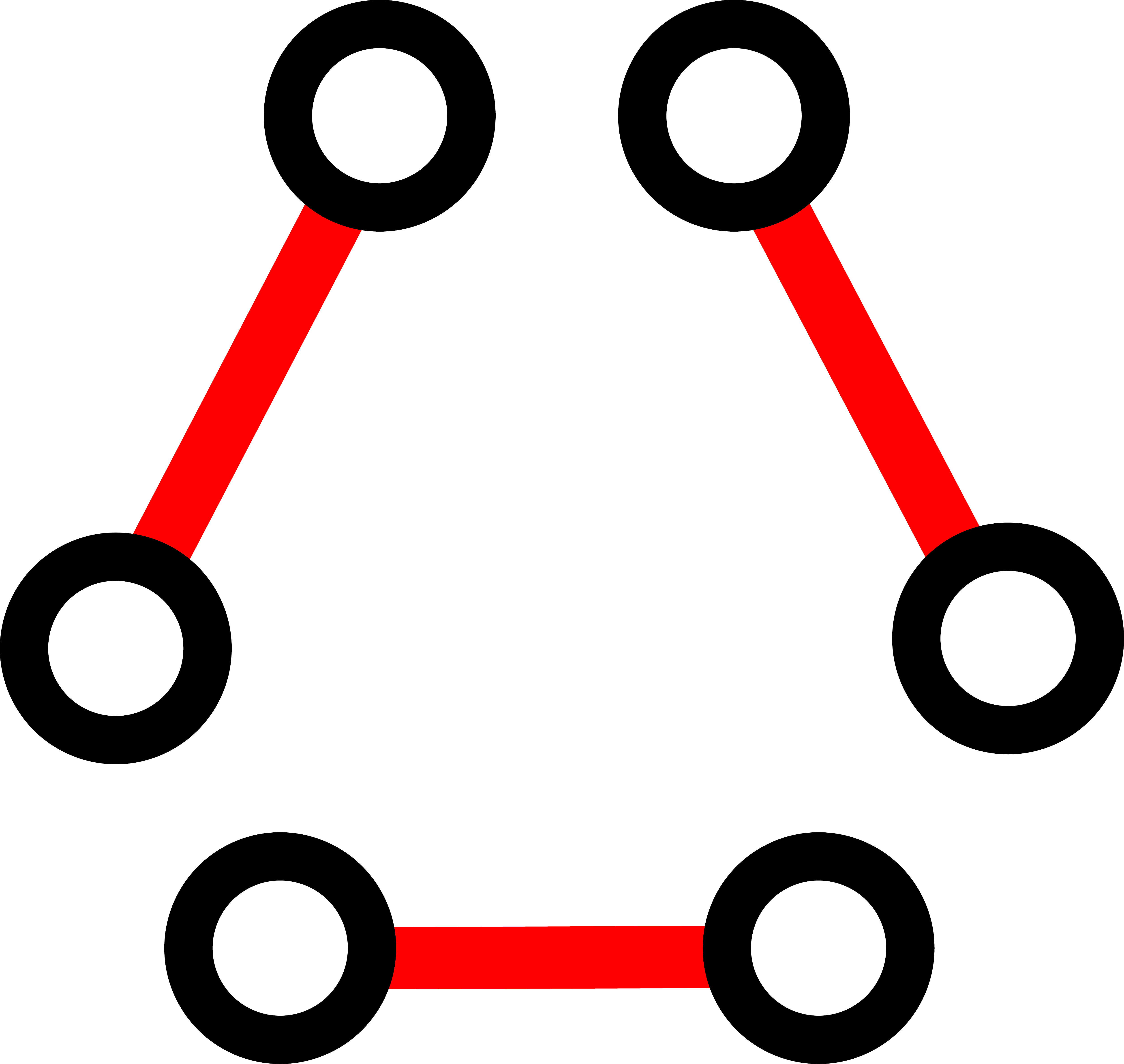}}
\newcommand*\ahex{\includegraphics[width=5mm]{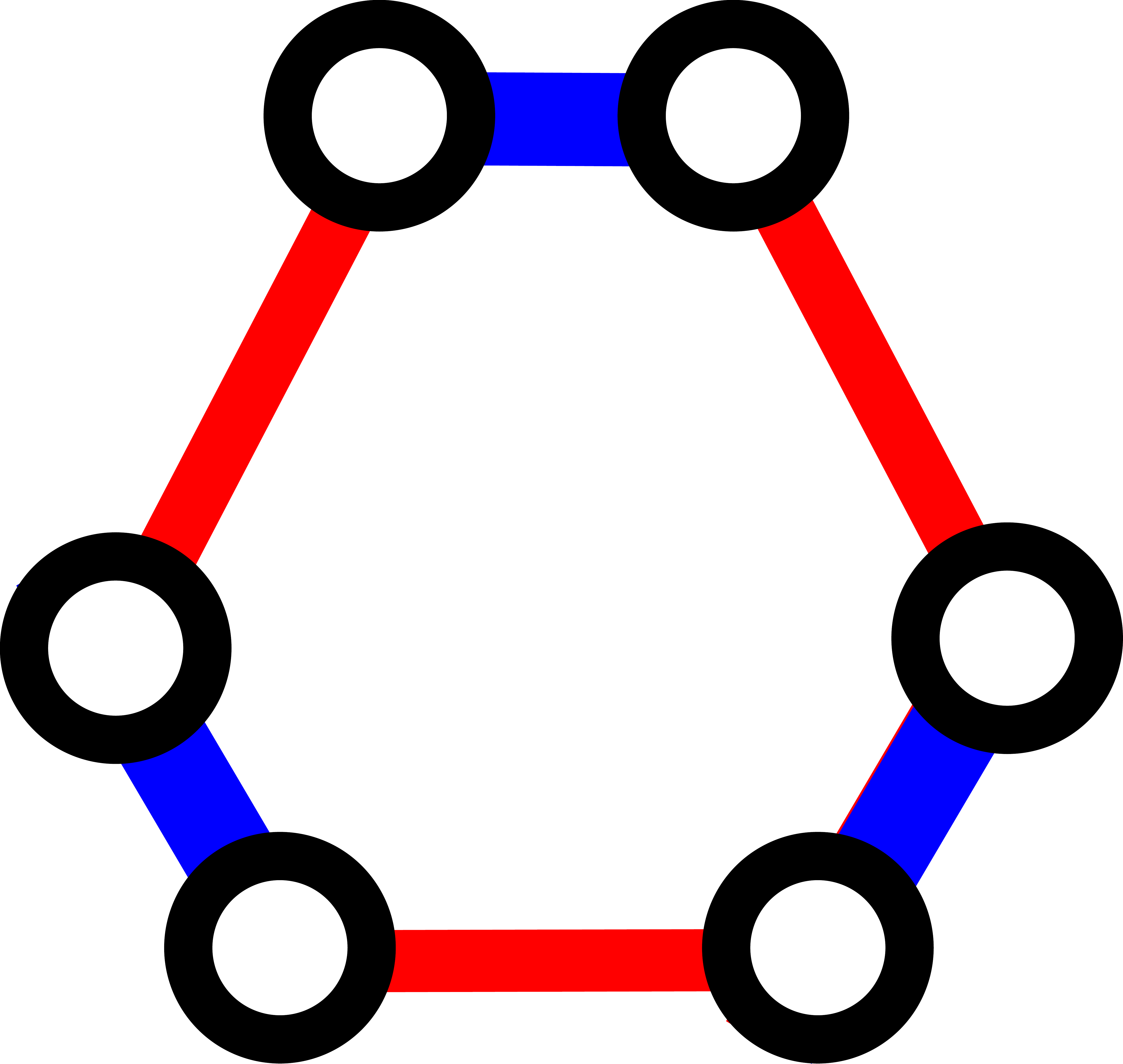}}
\newcommand*\uptri{\includegraphics[width=5mm]{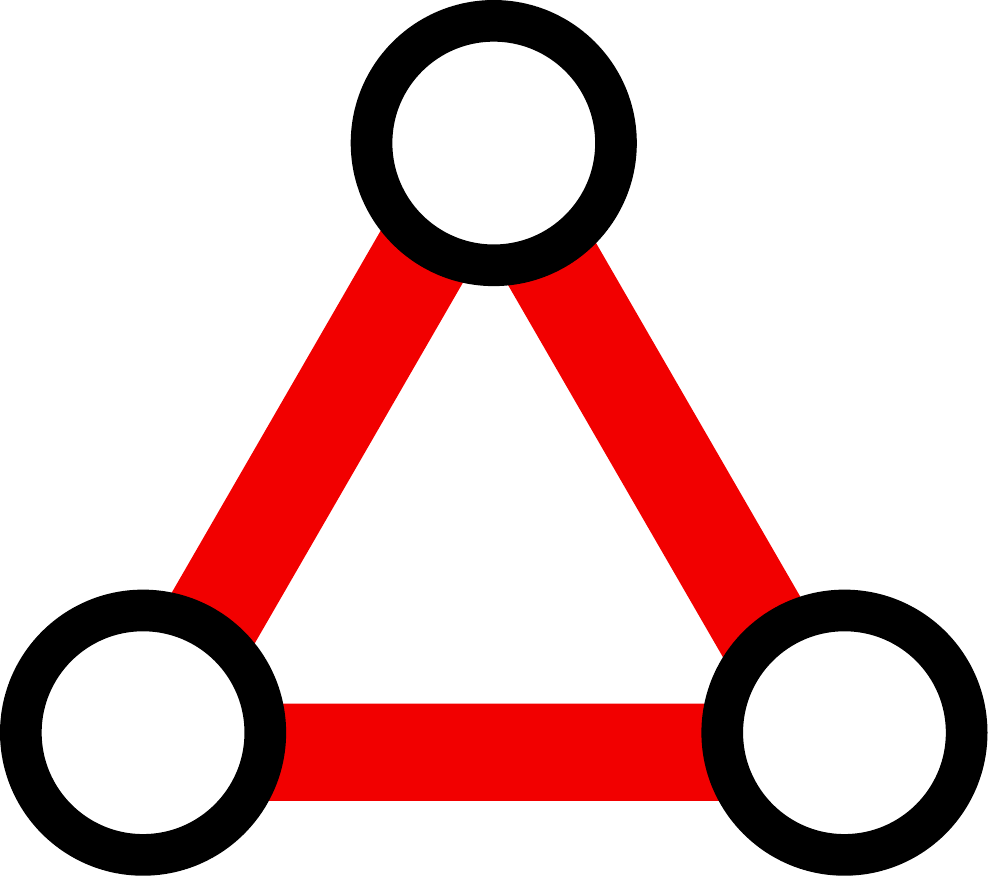}}
\newcommand*\dntri{\includegraphics[width=5mm]{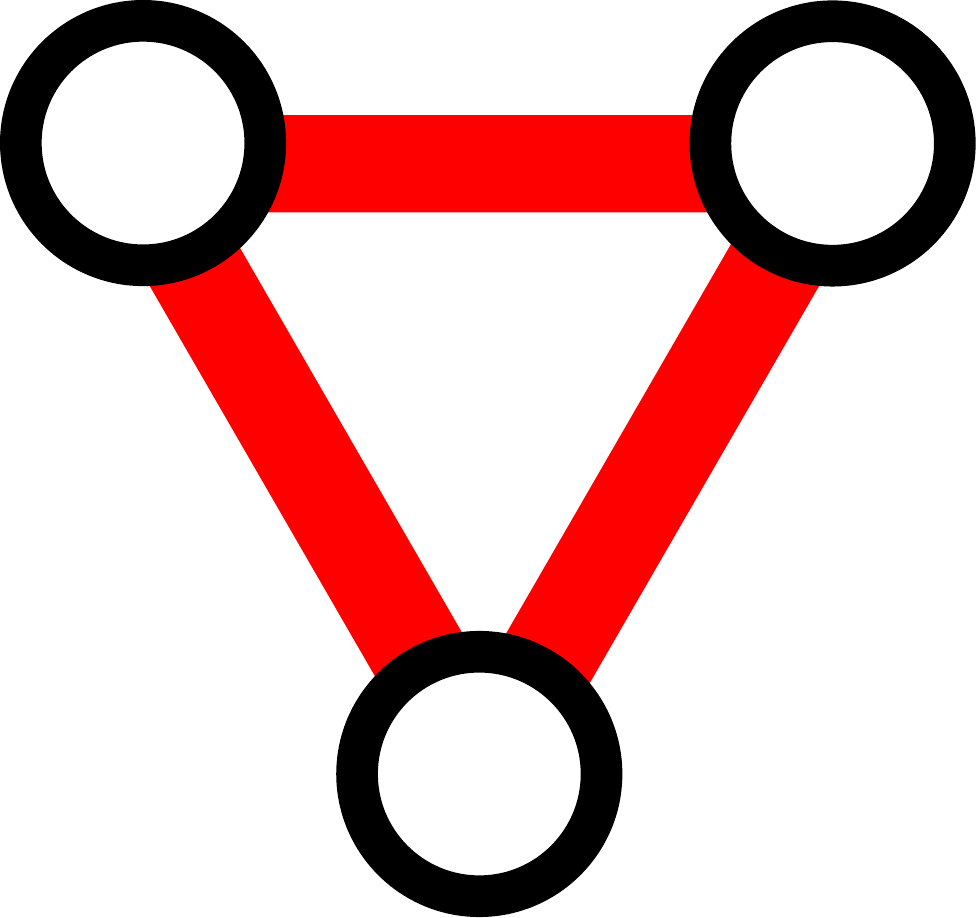}}
\makeatletter
\newcommand*{\rom}[1]{\expandafter\@slowromancap\romannumeral #1@}
\makeatother
\DeclareRobustCommand{\rchi}{{\mathpalette\irchi\relax}}
\newcommand{\irchi}[2]{\raisebox{\depth}{$#1\chi$}}
\newcommand{\cmmnt}[1]{}

\begin{document}

\title{Spontaneous dimerization and moment formation in the Hida model of spin-$1$ kagom\'e antiferromagnet}

\author{Pratyay Ghosh}
\affiliation{School of Physical Sciences, Jawaharlal Nehru University, New Delhi 110067, India}
\author{Brijesh Kumar}
\email{bkumar@mail.jnu.ac.in}
\affiliation{School of Physical Sciences, Jawaharlal Nehru University, New Delhi 110067, India}

\date{\today}

\begin{abstract}
The Hida model, defined on honeycomb lattice, is a spin-1/2 Heisenberg model of aniferromagnetic hexagons (with nearest-neighbor interaction, $J_A>0$) coupled via ferromagnetic bonds (with exchange interaction, $J_F <0$). It applies to the spin-gapped organic materials, \ce{m-MPYNN.X} (for $\mathrm{X}=$\ce{I}, \ce{BF4}, \ce{ClO4}), and  for $|J_F|\gg J_A$, it reduces to the spin-$1$ kagom\'e Heisenberg antiferromagnet (KHA). Motivated by the recent finding of trimerized singlet (TS) ground state for spin-1 KHA, we investigate the evolution of the ground state of Hida model from weak to strong $J_F/J_A$ using mean-field triplon analysis and Schwinger boson mean-field theory. Our triplon analysis of Hida model shows that its uniform hexagonal singlet (HS) ground state (for weak $J_F/J_A$) gives way to the dimerized hexagonal singlet (D-HS) ground state for $|J_F|/J_A \gtrsim 1.26$ (which for strong $J_F/J_A$ approaches the TS state). From the Schwinger boson calculations, we find that the evolution from the uniform HS phase for spin-1/2 Hida model to the TS phase for spin-1 KHA happens through two quantum phase transitions: 1) the spontaneous dimerization transition at $J_F/J_A \sim -0.28$ from the uniform HS to D-HS phase, and 2) the moment formation transition at $J_F/J_A \sim -1.46$, across which the pair of spin-1/2's on every FM bond begins to express as a bound moment that tends to spin-1 for large negative $J_F$'s. The TS ground state of spin-1 KHA is thus adiabatically connected to the D-HS ground state of the Hida model. Our calculations imply that the \ce{m-MPYNN.X} salts realize the D-HS phase at low temperatures, which can be ascertained through neutron diffraction.            
\end{abstract}

\maketitle

\section{Introduction}
The frustrated quantum spins at low temperatures are known to favor quantum-disordered phases  
such as the spin-liquid, valence-bond-solid, nematic, or dimer and plaquette ordered non-magnetic states ~\cite{Caspers89, Bose01, Misguich11, Penc11, Iqbal2016}. The kagom\'e antiferromagnet is one such example of a highly frustrated spin system of great current interest. The spin-$1/2$ kagome antiferromagnetic materials, such as \ce{Cu3Zn(OH)6Cl2} ~\cite{Olariu08,Mendels11} and \ce{BaCu3V2O8(OH)2}\cite{Okamoto2009}, show the absence of magnetic ordering down to very low temperatures, and are believed to realize some kind of a quantum spin-liquid state. But the true nature of the ground state of the spin-$1/2$ kagom\'e Heisenberg antiferromagnet (KHA) is a topic of ongoing theoretical debate~\cite{Zeng1990, Marston1991, Chalker1992, Mila1998, Li2012, Iqbal2015, Hering2017}.

The spin-$1$ kagom\'e antiferromagnetic case, realized for instance in \ce{m-MPYNN.X}~\cite{Awaga1994, Wada1997, Awaga1999, Kambe2004, Kambe2004a, Matsushita2010}, \ce{NaV6O11}~\cite{Kato2001} and \ce{KV3Ge2O9}~\cite{Hara2012,Takagi2017}, has also been a subject of recent investigations. Of these materials, perhaps the most studied is the family of organic salts \ce{m-MPYNN.X} (m-N-methylpyridinium~$\alpha$-nitronyl nitroxide) with $\mathrm{X}=$\ce{I}, \ce{BF4}, \ce{ClO4}, etc. These organic materials consist of strongly ferromagnetic spin-1/2 pairs coupled antiferromagnetically, which at low temperatures behave as spin-1 moments forming an antiferromagnetic kagom\'e lattice. The susceptibility measurements down to $35$ mK on these organic spin-1 kagom\'e compounds show a clear spin-gapped behavior and no magnetic ordering.  

Recent theoretical studies on spin-1 KHA clearly find a spin-gapped non-magnetic ground state, but with somewhat differing details~\cite{Hida2000, Gotze2011, Li2014.RAL, Liu2015, Picot2015, Changlani2015, Ghosh2016}. The two serious candidates for this ground state are the hexagonal singlet solid (HSS) state and the trimerized singlet (TS) state. Of these two, the TS state is favoured by most studies (based on tensor network algorithms, DMRG, triplon analysis etc.) as the ground state of the spin-1 KHA~\cite{Liu2015, Picot2015, Changlani2015, Ghosh2016}. Notably, the TS ground state spontaneously breaks the lattice symmetry by having more singlet weight on either all up triangles or all down triangles of the kaogm\'e lattice. It is therefore twofold degenerate. The spin-1 KHA, thus, presents us with an interesting case of spontaneous trimerization in a frustrated quantum antiferromagnet, beyond the spontaneous dimerization that we are so familiar with.

\begin{figure}[t]
\includegraphics[width=0.8\columnwidth]{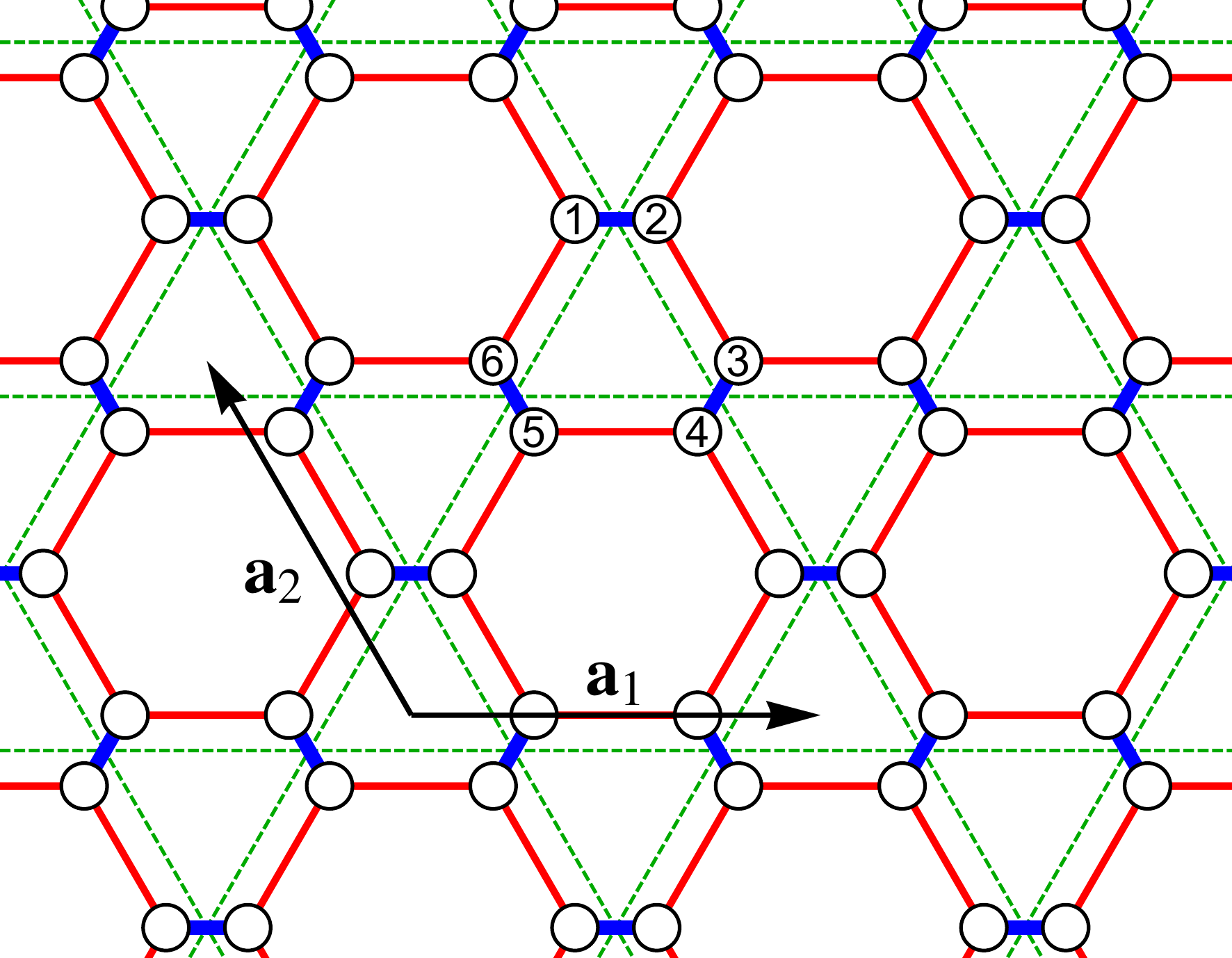}
\caption{\label{hidalattice}The Hida model with ferromagnetic Heisenberg exchange, $J_{F}$, shown as blue links, and the antiferromagnetic exchange interaction, $J_{A}$, shown in red color. The $\a_{1}=2\hat{x}$ and $\a_{2}=-\hat{x}+\sqrt{3}\hat{y}$ are two primitive vectors of the lattice. Moreover,  we define $\a_{3}=\a_{1}+\a_{2}$}
\end{figure}

The HSS state proposed by Hida \cite{Hida2000} breaks no lattice or spin symmetry. Its construction was inspired by the structure of \ce{m-MPYNN.BF4}, which is basically a spin-1/2 Heisenberg problem on a honeycomb lattice made of antiferromagnetic hexagons coupled ferromagnetically, as shown in Fig.~\ref{hidalattice}. Here, the intra-hexagon antiferromagnetic (AFM) interaction ($J_A>0$) is shown as red bonds, while the inter-hexagon ferromagnetic (FM) interaction ($J_F < 0$) is shown as blue bonds. This is how it was modeled by Hida~\cite{Hida2000}. Hence, we call it Hida model, which applies to the \ce{m-MPYNN.X} family. The HSS state can be constructed by first forming the direct product of the lowest energy singlet on every AFM (red) hexagon (as if they were independent of each other), and then symmetrizing the pair of spins on every FM (blue) bond. This is akin to the valence-bond solid state constructed by Affleck, Kennedy, Lieb and Tasaki for spin-1 chain~\cite{AKLT}. While the first step here tries to satisfy the AFM interaction locally on every hexagon, the second step forms a spin-1 out of two spin-1/2's on FM bonds. But as stated above, the HSS state turns out not to be the best choice for the ground state of spin-1 KHA, which is the large $|J_F|/J_A$ limit of the Hida model.

Without the symmetrization, however, the direct product of the hexagonal singlets is the exact ground state of the Hida model for $J_F=0$. Let us call this as the hexagonal singlet (HS) state, to distinguish it from the (symmetrized) HSS state. The question which interests us is that how the HS state (with uniform singlet amplitude) evolves to become the symmetry-breaking TS ground state with increasing $|J_F|/J_A$. Or, restating it differently, how the HS state surprisingly does not become the HSS state in the large $J_F$ limit? In this paper, we address this question by doing triplon analysis and Schwinger boson mean-field theory of the Hida model. The key findings from our Schwinger boson mean-field calculations are as follows: The uniform HS state first undergoes a spontaneous dimerization transition at $|J_{F}|/J_{A} = 0.28$, while the moments still behave as spin-1/2. That is, a small inter-hexagonal FM coupling induces dimerization of the singlet amplitude on the bonds of the AFM hexagons, as shown in Fig.~\ref{figHSS-DHSS-TS}(b). We call it the dimerized-HS (D-HS) state, which survives for all larger values of $|J_F|/J_A$. Then, around $|J_{F}|/J_{A} = 1.46$, a second transition occurs, under which the spin-1/2 moments on FM bonds begin to behave as bound pairs whose total moment per FM bond rapidly grows to spin-1. Thus, the D-HS state with fully formed spin-1 moments for $|J_{F}|/J_{A}\gg 1.46$ in Hida model is the TS state of spin-1 KHA. It is in qualitative agreement with the triplon analysis which also finds a transition from the uniform to dimerized HS state that smoothly approaches the TS state for large $|J_{F}|/J_{A}$. Furthermore, we discuss how to experimentally differentiate the HS from the D-HS state, and suggest that the low temperature phase of \ce{m-MPYNN.BF4} salt would be the D-HS (and not the HSS as originally suggested by Hida).

This paper is organized as follows. In Sec.~\ref{sec:model}, we describe the Hida model of quantum spin-1/2's. We study the evolution of its ground state with increasing $|J_F|/J_A$ using triplon mean-field theory (TMFT) in Sec.~\ref{sec:triplon_method} and Schwinger boson mean-field theory (SBMFT) in Sec.~\ref{sec:SB_method}. Both of these calculations produce mutually agreeable physics. We then conclude in Sec.~\ref{sec:summary}. 

\section{Hida Model}\label{sec:model}
The object of our study in this paper, the Hida model, is  a quantum spin-1/2 Heisenberg model on a honeycomb looking lattice that has the symmetries of a kagom\'e lattice due to a particular choice of the exchange interactions (motivated by the organic salt, \ce{m-MPYNN.BF4}). As shown in Fig.~\ref{hidalattice}, it can be best described as a model of antiferromagnetic hexagons, coupled ferromagnetically. Here, the hexagons with nearest-neighbor AFM interaction, $J_A>0$, are shown in red color, and the thick blue bonds depict the inter-hexagonal FM interaction, $J_F<0$. The dotted green lines, joining the centers of the FM bonds, are drawn to indicate the underlying kagom\'e lattice. The unit-cell of the Hida model has six spins, as labeled in the figure. The Hamiltonian of the Hida model is given below.
\begin{equation}\label{eq:Hamil}
\hat{H} = J_{F} \sum_{{\langle i,j \rangle}}^{\fer} \vec{S}_{i} \cdot \vec{S}_{j}
+J_{A} \sum_{{\langle i,j \rangle}}^{\afer} \vec{S}_{i} \cdot \vec{S}_{j}
\end{equation}
Every $\vec{S}_{i}$ here is a spin-$1/2$ operator. Some experimental values of $J_F$ and $J_A$, estimated from the susceptibility measurements on the \ce{m-MPYNN.X} family of compounds~\cite{Awaga1994, Wada1997}, are presented in Table~\ref{tab:mpynn}.   
\begin{table}[b]
\centering
\caption{The exchange interactions for $\mathrm{m-MPYNN^{+}\cdot X^{+}} \cdot \frac{1}{3} \mathrm{(acetone)}$ from Refs.~\onlinecite{Awaga1994, Wada1997}.}
\label{tab:mpynn}
\begin{tabular}{|l|l|l|l|l|l|}
\hline
$\mathrm{X}$ & $\mathrm{I}$ & $\mathrm{BF_{4}}$ & $\mathrm{(BF_{4})_{0.72}I_{0.28}}$ & $\mathrm{ClO_{4}}$  \\
\hline
$J_A$                  & $1.6$ K & $3.11$ K &$1.20$ K    & $0.19$ K      \\
\hline
$J_F$               & $-10.2$ K  & $-23.26$ K &$-11.3$ K   &  $-10.5$ K         \\ 
\hline
$J_{F}/J_{A}$              & $-6.375$  & $-7.479$  &$-9.416$   & $-55.263$       \\ 
\hline     
\end{tabular}
\end{table} 
In the limit $|J_{F}|/J_A \rightarrow \infty$, the Hida model exactly becomes the spin-$1$ KHA model, $\mathcal{H}_{KHA}=\widetilde{J}_{A}\sum_{\langle i,j \rangle}{\bf S}_{i}\cdot{\bf S}_{j} $, with the nearest-neighbor interaction, $\widetilde{J}_{A}=J_{A}/4$~\cite{Hida2000}. 
 
The Hida model for $J_F=0$ is a model of independent hexagons with a trivial ground state in which every AFM hexagon is in its lowest energy singlet state. How this uniform HS (hexagonal singlet) ground state changes with $J_F$, and eventually becomes the TS ground state for large enough $J_F$, is the question that we address in the next two sections. By doing triplon mean-field theory (TMFT), we first compare the energies of the candidate states to see their relative tendencies as a function of $|J_F|/J_A$. Next, we do a Schwinger boson mean-field theory (SBMFT) of the Hida model, which gives us a clear understanding of the transition from the HS to the TS phase in the ground state. 
\section{Triplon Mean-Field Theory}\label{sec:triplon_method}
\begin{figure}
\includegraphics[width=0.7\columnwidth]{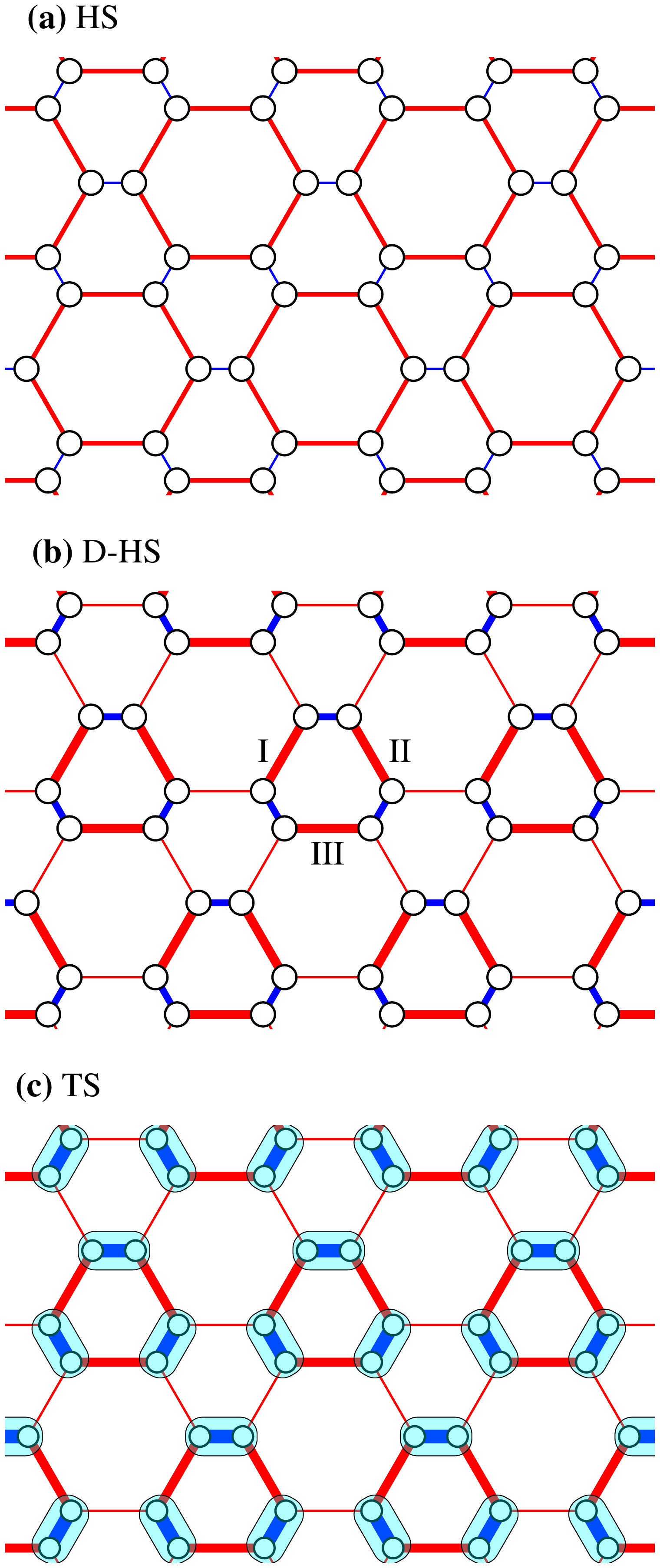}
\caption{\label{figHSS-DHSS-TS}
Three possible ground states of the Hida model. (a) The hexagonal singlet (HS) state with uniform singlet amplitude per bond on every antiferromagnetic (red) hexagon. (b) The D-HS state with dimerized AFM bonds (shown as thick and thin red bonds). (c) Trimerized Singlet (TS) state.
} 
\end{figure}

The triplon mean-field theory is a low-energy bosonic theory of the triplet fluctuations for a given non-magnetic quantum state. In our previous work, we did such a theory of the TS state for spin-1 KHA~\cite{Ghosh2016}. This approach provides a simple means to study the renormalization and the stability of a reference state against its low-energy quantum fluctuations. For Hida model, we identify three singlet states plausible to be the ground state for different ranges of $|J_F|/J_A$. These are shown in Fig.~\ref{figHSS-DHSS-TS}. 

The state depicted in Fig.~\ref{figHSS-DHSS-TS}(a) is the HS state, in which all the AFM (red) hexagons form the singlet (with uniform amplitude per bond). It preserves all the symmetries of the underlying lattice, and is expected to be the ground state for small $|J_F|/J_A$. For $J_F=0$, it is anyway the exact ground state. The state shown in Fig.~\ref{figHSS-DHSS-TS}(c) is the TS state (analogous to the spin-1 KHA), in which all up-oriented, $\afup$, (or all down-oriented, $\afdn$) hexagons with alternate red and blue bonds form the singlet. This state breaks the lattice symmetry exactly in the same way as the trimerized singlet state of the spin-1 KHA, and is expected to be the ground state when $|J_F|\gg J_A$. Upon a careful observation, we realize that this TS state can also be viewed as the HS state with dimerized singlet amplitudes (with alternate strong and weak red bonds in every antiferromagnetic hexagon). It presents us with an interesting third state shown in Fig.~\ref{figHSS-DHSS-TS}(b), which we call as the dimerized-HS  or D-HS state. In this state, we do not bother about forming the lowest-energy singlet on a hexagon as a whole. Instead, we only form the dimer singlets on the alternate AFM bonds, as in Fig.~\ref{figHSS-DHSS-TS}(b). Using TMFT, we compute the energies of these three states as a function of $|J_F|/J_A$, and see how they compete to be the ground state.
 
For doing TMFT, we first derive the representation of the spin-1/2 operators in terms of the lowest energy singlet and triplet eigenstates of the individual elementary blocks on which the candidate state forms the singlet. For the HS state, the elementary blocks are the AFM hexagons, $\afer$; for the TS state, these blocks are, say, the up-oriented hexagons with alternate AFM-FM bonds, $\ahex$; and for the D-HS state, only the three AFM bonds on up-oriented AFM-FM hexagons, $\afup$, are individually treated as the elementary blocks. We find the eigenstates and eigenvalues of the corresponding block Heisenberg Hamiltonians separately for the three cases. Of these, we keep only the lowest lying singlet and the triplets immediately above it, and ignore the rest of the higher energy states, as we are interested in the minimal low-energy description of the system with respect to the three candidate states. Then, in this reduced basis, $\{|b_{k}\rangle\}$, we write the basic spin-1/2 operators on the hexagons as \( S_{j,\alpha}=\sum_{k,l}\mathcal{M}_{j,\alpha}^{k,l}|b_{l}\rangle\langle b_{k}| \), where $\mathcal{M}_{j,\alpha}^{k,l}=\langle b_{k}|S_{j,\alpha}|b_{l}\rangle$, $j=1$ to 6 is the spin label (as in Fig.~\ref{hidalattice}), $\alpha=x,y,z$ are the three components of the spin operators.

For further simplification, we approximate the singlet state on every elementary block by a mean singlet amplitude, $\bar{s}$. We treat the triplet states by associating to them the bosonic triplon operators, and keep in the representation of the spin-1/2 operators only those triplon terms that couple to $\bar{s}$. This latter approximation amounts to neglecting the triplon-triplon interaction in the full Hamiltonian. We then rewrite the full Hamiltonian, $\hat{H}$ of Eq.~\eqref{eq:Hamil}, in this triplon representation of the spin-1/2 operators. The constraint on the total number of bosons is satisfied via a mean Lagrange multiplier, $\lambda$. These steps lead to a Hamiltonian which is bilinear in the triplon operators, and describes the effective low-energy triplon dynamics of the Hida model. Below we formulate the TMFT separately for the HS, D-HS and TS states.

\subsection{Hexagonal Singlet (HS) State}
Using the convention given in Fig.~\ref{hidalattice}, we write the spin-$1/2$ block Hamiltonian of a single AFM hexagon as  
\begin{eqnarray}
\mathcal{H}_{\afer} & = &
J_{A} \left\{\vec{S}_{3}(\r)\cdot \left[\vec{S}_{2}(\r)+\vec{S}_{6}(\r+\a_{1})\right]\right.\nonumber\\
 && +\left.\vec{S}_{1}(\r+\a_{1}) \cdot \left[\vec{S}_{6}(\r+\a_{1})+\vec{S}_{4}(\r+\a_{3})\right]\right.\nonumber\\
 && +\left.\vec{S}_{5}(\r+\a_{3}) \cdot \left[\vec{S}_{4}(\r+\a_{3})+ \vec{S}_{2}(\r)\right]\right\}.
\end{eqnarray}

We use the sixfold rotational symmetry associated with an isolated AFM hexagon to write its eigenstates. The sixfold rotation operator, $R$, rotates the hexagon by $60^{\circ}$, and its eigenvalues are: $\lambda_{R}=\pm 1$, $\pm \omega$, $\pm \omega^{2}$. Its lowest energy eigenstate is a unique spin-singlet with energy $ E_{s}=-\frac{J_A}{2}\left(2+\sqrt{13}\right)$ and $\lambda_R=1$. The next higher energy eigenstate is a triplet of $m=-1,0,1$ (total $S_{z}$) with energy $E_{t}=-\frac{J_A}{2}\left(2+\sqrt{5}\right)$ and $\lambda_{R}=-1$. To keep it minimal, we neglect the higher energy eigenstates of ${\afer}$, which reduces the local Hilbert space to one singlet $|s\rangle$ and a triplet, $\{|t_{m}\rangle\}$. 

Similar to the bond-operator formalism \cite{Sachdev1990, Kumar2010, Kumar2008}, we write the spin-$1/2$ operators on an AFM hexagon in the reduced basis in terms of the hexagonal singlet and triplet operators, respectively $\shat^{\dagger}$ and $\that^{\dagger}_{m}$, acting in a bosonic Fock space. The projection of the infinite dimensional Fock space onto the $4$-dimensional reduced Hilbert space spanned by $|s\rangle$ and $|t_{m}\rangle$ is done by the constraint, $ \shat^{\dagger}\shat+\sum_{m}\that^{\dagger}_{m}\that_{m}=1$, on the number of these bosons. In the reduced space, the local hexagonal block Hamiltonian can now be written as: 
\begin{equation} \label{eq:unihex}
 \mathcal{H}_{\afer}\approx E_{s}\shat^{\dagger}\shat+E_{t}\sum_{m}\that_{m}^{\dagger}\that_{m}.
\end{equation}
The spin-1/2 operators on an AFM hexagon can approximately be represented as: 
\begin{equation} \label{eq:spinHSS}
 S_{i, \alpha}  \approx\mathcal{C}_{i}\sbar\Qhat_{\alpha}
\end{equation}
where for $i^\prime=1,2$ and $3$,
\begin{eqnarray}
 \mathcal{C}_{2 i^\prime-1}&=&-\sqrt{2}\langle t_{0}|S_{1}^{z}|s\rangle=2\langle t_{1}|S_{1}^{+}|s\rangle=2\langle t_{\onebar}|S_{1}^{-}|s\rangle\nonumber\\
 &=&\frac{7\sqrt{5}+7\sqrt{13}+2\sqrt{65}+26}{6\sqrt{6\left(65+18\sqrt{13}\right)}}  =-\mathcal{C}_{2 i^\prime}=-\mathcal{C} 
\end{eqnarray}
Moreover, the ``coordinate" operator 
\begin{equation}
 \Qhat_{\alpha}=\frac{1}{\sqrt{2}}\left(\that^{\dagger}_{\alpha}+\that_{\alpha}\right) 
\end{equation}
for
$\that^\dag_{x} = \frac{1}{\sqrt{2}}\left(\that^\dag_{\onebar}-\that^\dag_{1}\right)$,  
$ \that^\dag_{y} = -\frac{i}{\sqrt{2}}\left(\that^\dag_{\onebar}+\that^\dag_{1}\right) $ and 
$ \that^\dag_{z} = -\that^\dag_{0} $.
The ``momentum'' operator conjugate to $\Qhat_{\alpha}$ is defined as $\Phat_{\alpha}=\frac{i}{\sqrt{2}}\left(\that^{\dagger}_{\alpha}-\that_{\alpha}\right)$. Here, we have treated the singlet operator, $\shat$, as mean-field, $\sbar$. Through $\sbar$, which measures the mean singlet amplitude per AFM hexagon, we describe in mean-field approximation the HS phase of the Hida model. For a general discussion on triplon mean-field theory, please take a look at the Refs.~\cite{Sachdev1990, Kumar2010, Kumar2008, Ghosh2016}.  

By using Eqs.~\eqref{eq:unihex} and~\eqref{eq:spinHSS}, we turn the Hida model of Eq.~\eqref{eq:Hamil} into an effective theory of triplons with respect to the HS state. The effective triplon Hamiltonian has the following form in momentum space.
\begin{eqnarray} \label{eq:hamilHSS}
 \hat{H}_{t}^{HS}=e_{0}N_{uc}&+&\frac{1}{2}\sum_{\k}\sum_{\alpha=x,y,z}\left[\lambda\Phat_{\alpha}^{\dagger}\left(\k\right)\Phat_{\alpha}\left(\k\right)\right.\nonumber\\
 &+&\left.\left(\lambda- 2J_{F}\mathcal{C}^{2}\sbar^{2}\rchi_{\k}\right)\Qhat_{\alpha}^{\dagger}\left(\k\right)\Qhat_{\alpha}\left(\k\right)\right]
\end{eqnarray}
Here, $N_{uc}$ is the total number of hexagonal unit-cells in the lattice, $e_{0}=\left(E_{s}-E_{t}\right)\sbar^{2}+\lambda\sbar^{2}-\frac{5}{2}\lambda+E_{t}$, and $\lambda$ is the Lagrange multiplier that is introduced to satisfy the local constraint, $\shat^{\dagger}\shat+\sum_{\alpha}\that_{\alpha}^{\dagger}\that_{\alpha}=1$, on average. The operator $\Qhat_{\alpha}(\k)$ is the Fourier transform of $\Qhat_{\alpha}(\r)$, where $\r$ denotes the position vector of the hexagonal unit cell of the lattice, 
and $\k$ is the lattice-momentum vector in the first Brillouin zone of the corresponding reciprocal lattice. Likewise, $\Phat_{\alpha}(\r) = \frac{1}{\sqrt{N_{uc}}}\sum_\k e^{i\k\cdot\r}\Phat_{\alpha}(\k)$. Since $\Qhat_{\alpha}(\r)$ and $\Phat_{\alpha}(\r)$ are Hermitian, therefore, $\Qhat^{\dagger}_{\alpha}(\k) = \Qhat_{\alpha}(-\k)$ and $\Phat^{\dagger}_{\alpha}(\k) = \Phat_{\alpha}(-\k)$. Moreover, $[\Qhat_{\alpha}(\k),\Phat_{\alpha^\prime}(\k^\prime)] = i\delta_{\alpha\alpha^\prime}\delta_{\k+\k^\prime=0}$, while the $\Qhat_{\alpha}(\k)$'s commute amongst themselves and the same for $\Phat_{\alpha}(\k)$'s. In Eq.~\eqref{eq:hamilHSS},
\begin{equation}
 \rchi_{\k}=\cos{\k\cdot\bf{a}_{1}}+\cos{\k\cdot\bf{a}_{2}}+\cos{\k\cdot\bf{a}_{3}}, 
\end{equation}
where $\bf{a}_{1}$, $\bf{a}_{2}$ and $\bf{a}_{3}$ are as defined in Fig.~\ref{hidalattice}.

The Hamiltonian in Eq.~\ref{eq:hamilHSS} is essentially a problem of three coupled harmonic oscillators. In the diagonal form, the $\hat{H}_{t}^{HS}$ can be written as follows.
\begin{equation}
 \hat{H}_{t}^{HS}=e_{0}N_{uc}+\sum_{k}\sum_{\alpha=x,y,z}\omega_{\alpha,\k}\left[\hat{\gamma}^{\dagger}_{\alpha}\left(\k\right)\hat{\gamma}_{\alpha}\left(\k\right)+\frac{1}{2}\right]
\end{equation}
Here, $\hat{\gamma}_{\alpha}\left(\k\right)=\sqrt{\frac{\omega_{\alpha,\k}}{2\lambda}}\Qhat_{\alpha}\left(\k\right)+i\sqrt{\frac{\lambda}{2\omega_{\alpha,\k}}}\Phat_{\alpha}\left(\k\right)$ are the renormalized triplon operators, and
\begin{equation}
 \omega_{\alpha,\k}=\sqrt{\lambda\left(\lambda-\sbar^{2}\xi_{\alpha,\k}\right)}
\end{equation}
are the triplon energy dispersions with $\xi_{\alpha,\k}=2J_{F}\mathcal{C}^{2}\rchi_{\k}$.
The ground state energy per unit-cell is
\begin{equation}
 e^{HS}_{g}=e_{0}+\frac{1}{2N_{uc}}\sum_{k}\sum_{\alpha=x,y,z}\omega_{\alpha,\k}.
\end{equation}
It is a function of two unknown mean-field parameters, $\lambda$ and $\sbar^2$. We determine them by minimizing $e_g$. The $\partial_\lambda e_g  =0$ and $\partial_{\sbar^2} e_g =0$ give us the following mean-field equations, whose self-consistent solution gives the physical values of $\lambda$ and $\sbar^{2}$.
\begin{subequations} \label{sc-HSS}
 \begin{align}
  \sbar^2 &= \frac{5}{2}-\frac{1}{4N_{uc}}\sum_{\k}\sum_{\alpha=x,y,z}\frac{2\lambda-\sbar^{2}\xi_{\alpha,\k}}{\omega_{\alpha,\k}} \label{eq:sbar} \\
   \lambda &=\left(E_{t}-E_{s}\right)+\frac{\lambda}{4N_{uc}}\sum_{\k}\sum_{\alpha=x,y,z}\frac{\lambda\xi_{\alpha,\k}}{\omega_{\alpha,\k}} \label{eq:lam}
 \end{align}
\end{subequations}

This formalism would present us with two physical solutions, viz., the gapped or the gapless triplons. When the minimum of the lowest of these dispersions in the Brillouin zone is strictly greater than zero, it means there is an energy gap that protects the HS ground state against triplon excitations. We surely expect this to happen when $J_F$ is near about zero.

\subsection{Dimerized Hexagonal Singlet (D-HS) State}
To describe the D-HS state of the Hida model, we start with the Heisenberg model for only the AFM bonds of an up-oriented hexagon with alternate AFM-FM bonds.
\begin{eqnarray}\label{eq:hamilhexDHSS}
\mathcal{H}_{\afup} &=& J_{A}\sum_{i=1,2,3}\vec{S}_{2i}\cdot\vec{S}_{2i+1}\\
&&  \hspace{2cm} (\mbox{with} ~\vec{S}_{7}=\vec{S}_{1}) \nonumber
\end{eqnarray}
Using the well-known bond-operator representation~\cite{Sachdev1990, Kumar2010, Kumar2008}, we write the spin-1/2 operators on these three AFM bonds in terms of the singlet and triplet bosonic operators. 
\begin{eqnarray}
 S_{2i}^{\alpha}&\approx & \frac{\bar{s}}{2}\left(\that_{i,\alpha}+\that_{i,\alpha}^{\dagger}\right)\nonumber\\
 S_{2i+1}^{\alpha}& \approx & -\frac{\bar{s}}{2}\left(\that_{i,\alpha}+\that_{i,\alpha}^{\dagger}\right)
\end{eqnarray}
Here, the $\bar{s}$ is the mean singlet amplitude on these AFM bonds. The block Hamiltonian of Eq.~\eqref{eq:hamilhexDHSS} now reads as: 
\begin{equation}\label{eq:hamilhexDHSS1}
 \mathcal{H}_{\afdhss}=-\frac{9}{4}J_{A}\sum_{i=\rom{1}}^{\rom{3}}\shat_{i}^{\dagger}\shat_{i}+\frac{3}{4}J_{A}\sum_{\alpha=x,y,z}\sum_{i=\rom{1}}^{\rom{3}}\that_{i,\alpha}^{\dagger}\that_{i,\alpha}
\end{equation}
The roman numerals indicate the three AFM bonds [see Fig.~\ref{figHSS-DHSS-TS}(b)]. The three constraints to be satisfied on these AFM bond are:
\begin{equation}\label{eq:consDHSS}
 \shat_{i}^{\dagger}\shat_{i}+\sum_{\alpha}\that_{i,\alpha}^{\dagger}\that_{i,\alpha}=1\ \ \forall i=\rom{1},\rom{2},\rom{3}.
\end{equation}
With the ``coordinate" operators defined as
\begin{equation}
 \Qhat_{i,\alpha}=\frac{1}{\sqrt{2}}\left(\that_{i,\alpha}^{\dagger}+\that_{i,\alpha}\right),
\end{equation}
the conjugate ``momentum'' operators as
\begin{equation}
 \Phat_{i,\alpha}=\frac{i}{\sqrt{2}}\left(\that_{i,\alpha}^{\dagger}-\that_{i,\alpha}\right),
\end{equation}
and their commutation properties as: $\left[\Qhat_{j,\beta},\Phat_{i,\alpha}\right]=i\delta_{ji}\delta_{\beta\alpha}$ and $\Phat_{i,\alpha}^{2}+\Qhat_{i,\alpha}^{2}=2\that_{i,\alpha}^{\dagger}\that_{i,\alpha}+1$, 
the bond-operator representation of the spins on an up-oriented AFM-FM hexagon can be written as follows.
\begin{subequations}\label{eq:spinDHSS}
\begin{eqnarray}\label{eq:spinDHSS}
S_{2,\alpha}& = &\frac{1}{\sqrt{2}}\bar{s}\Qhat_{\rom{2},\alpha} = -S_{3,\alpha} \\
S_{4,\alpha}& = &\frac{1}{\sqrt{2}}\bar{s}\Qhat_{\rom{3},\alpha} = -S_{5,\alpha} \\
S_{6,\alpha}&=&\frac{1}{\sqrt{2}}\bar{s}\Qhat_{\rom{1},\alpha} = -S_{1,\alpha} 
\end{eqnarray}
\end{subequations}

In this representation, the Hida model with reference to the D-HS state reads as: 
\begin{eqnarray} \label{hmfdhss}
 \hat{H}_{t}^{D\mbox{-}HS} & = & \frac{1}{2}\sum_{\k}\sum_{\alpha}\left[ \lambda \bf{\hat{P}}_{\alpha}^{\dagger}\left(\k\right) \bf{\hat{P}}_{\alpha}\left(\k\right) + \bf{\hat{Q}}_{\alpha}^{\dagger}\left(\k\right)\mathcal{V}_{\k}^{Q}\bf{\hat{Q}}_{\alpha}\left(\k\right)\right] \nonumber \\
 & & + \, e_{0}\,N_{uc}, 
\end{eqnarray}
where $e_{0}=-3J_{A}\bar{s}^{2}+\frac{3}{4}J_{A}+3\lambda\bar{s}^{2}-\frac{15}{2}\lambda$, ${\bf{\hat{Q}}}_{\alpha}^{\dagger}\left(\k\right)=\left[\Qhat_{\rom{1},\alpha}^{\dagger}\left(\k\right)\ \Qhat_{\rom{2},\alpha}^{\dagger}\left(\k\right)\ \Qhat_{\rom{3},\alpha}^{\dagger}\left(\k\right)\right]$ and ${\bf{\hat{P}}}_{\alpha}^{\dagger}\left(\k\right)=\left[\Phat_{\rom{1},\alpha}^{\dagger}\left(\k\right)\ \Phat_{\rom{2},\alpha}^{\dagger}\left(\k\right)\ \Phat_{\rom{3},\alpha}^{\dagger}\left(\k\right)\right]$. The Lagrange multiplier, $\lambda$, is introduced to satisfy the constraints in Eq. \ref{eq:consDHSS} on average. The Fourier transform of the operators is given as follows.
\begin{eqnarray}
\Qhat_{i,\alpha}\left({\r}\right)&=&\frac{1}{\sqrt{N_{uc}}}\sum_{\k}e^{i\k\cdot\r}\Qhat_{i,\alpha}\left(\k\right)\\
\Phat_{i,\alpha}\left({\r}\right)&=&\frac{1}{\sqrt{N_{uc}}}\sum_{\k}e^{i\k\cdot\r}\Phat_{i,\alpha}\left(\k\right)
\end{eqnarray}
In Eq.~\ref{hmfdhss}, 
\begin{eqnarray}
\mathcal{V}_{\k}^{Q}=& \lambda & \mathbb{I}_{3\times3}-\bar{s}^{2}J_{F}\left[ \begin{array}{ccc}
 0 & 1 & 1 \\
1 &  0 & 1\\
1& 1 & 0 \end{array}\right]\nonumber\\
&-&\bar{s}^{2}J_{A}\left[ \begin{array}{ccc}
 0 & e^{i\k\cdot\a_{2}} & e^{i\k\cdot\a_{3}}   \\
e^{-i\k\cdot\a_{2}} &  0 & e^{i\k\cdot\a_{1}}\\
e^{-i\k\cdot\a_{3}}& e^{-i\k\cdot\a_{1}} & 0 \end{array}\right]
\end{eqnarray}
is a Hermitian matrix. Here, $\mathbb{I}_{3\times 3}$ denotes the three-dimensional identity matrix.

The $\hat{H}_t^{D\mbox{-}HS}$ is a problem of three coupled oscillators for each $\alpha$ separately. Its ground state energy per unit cell is found to be
\begin{equation}
e^{D\mbox{-}HS}_{g}=e_{0}+\frac{1}{2N_{uc}}\sum_{\k}\sum_{\alpha}\sum_{i=\rom{1}}^{\rom{3}}E_{i}\left(\k\right),                                                                                                                                                                                                                                                                                                    \end{equation}
with $E_{i}\left(\k\right)=\sqrt{\lambda\left(\lambda-2\bar{s}^{2}\xi_{i}\left(\k\right)\right)}$. Here, 
\begin{align}
\xi_{i}(\k) =& ~ \left\{ \sqrt[3]{\rchi_{\k}+\sqrt{\rchi_{\k}^{2}+(\zeta_{\k}/3)^{3}}} \right. \nonumber  \\
& ~~~ + \left. \sqrt[3]{\rchi_{\k}-\sqrt{\rchi_{\k}^{2}+(\zeta_{\k}/3)^{3}}} ~ \right\}_i   \label{eq:cubsol}
\end{align}
with 
\begin{align*}
\rchi_{\k} & =\Re\left[\left(J_{A}+J_{F}e^{i\k.\a_{1}}\right)\left(J_{A}+J_{F}e^{i\k\cdot\a_{2}}\right)\right.\\
& \hspace{100pt} \left.\left(J_{A}+J_{F}e^{-i\k\cdot\a_{3}}\right)\right]
\end{align*} 
and 
\begin{align*}
\zeta_{\k} = & ~ 2J_{F}J_{A}\left(\cos\k\cdot\a_{1}+\cos\k\cdot\a_{2}+\cos\k\cdot\a_{3}\right) \\
& + 3\left(J_{F}^{2}+J_{A}^{2}\right).
\end{align*} 
Here, $ \xi_{i}(\k)$'s are the three roots of the cubic equation: $x^{3} - \zeta_{\k}x - 2\rchi_{\k} =0$. See footnote~\footnote{Cardano's formula: The roots of the cubic equation $x^{3}+px+q=0$ (for $p,q\in\mathbb{C}$) are given by $x=\alpha+\beta$ with, $\alpha=\sqrt[3]{-\frac{q}{2}+\sqrt{\frac{q^{2}}{4}+(\frac{p}{3})^{3}}}$ and $\beta=\sqrt[3]{-\frac{q}{2}-\sqrt{\frac{q^{2}}{4}+(\frac{p}{3})^{3}}}$. The only three valid combinations of $\alpha$ and $\beta$ are those for which $\alpha\beta=-p/3$ holds true.} on its solution of the allowed $\xi_i(\k)$'s as given in Eq.~\eqref{eq:cubsol}. 
By minimizing $e^{\mbox{\footnotesize $D$-$HS$}}_g$, we get the following two equations which can be solved self-consistently for $\lambda$ and $\sbar^{2}$. 
\begin{subequations} \label{sc-DHSS}
\begin{align}
 & \sbar^{2} =\frac{5}{2}-\frac{1}{6N_{uc}}\sum_{\k}\sum_{\alpha}\sum_{i=\rom{1}}^{\rom{3}}\frac{\lambda-\sbar^{2}\xi_{i}\left(\k\right)}{E_{i}\left(\k\right)}\\
 & \lambda =J_{A}+\frac{\lambda}{6N_{uc}}\sum_{\k}\sum_{\alpha}\sum_{i=\rom{1}}^{\rom{3}}\frac{\xi_{i}\left(\k\right)}{E_{i}\left(\k\right)}
  \end{align}
\end{subequations}
 
\subsection{Trimerized Singlet (TS) State}
In the TS state, we choose as reference the up-oriented hexagons (with alternate AFM-FM bonds) where the singlets are formed. The block Hamiltonian for each such up hexagon can be written (with $\vec{S}_{7}=\vec{S}_{1}$) as:
\begin{equation}
 \mathcal{H}_{\ahex}=J_{F}\sum_{i=1,2,3}^{\fer}\vec{S}_{2i-1}\cdot\vec{S}_{2i}+J_{A}\sum_{i=1,2,3}^{\afup}\vec{S}_{2i}\cdot\vec{S}_{2i+1}.
\end{equation}

Unlike the AFM hexagon, this only has threefold rotational symmetry (as for a triangular unit cell in kagom\'e lattice). Using this threefold rotation symmetry, we write the eigenstates of the up AFM-FM hexagon in the basis of the rotation operator, which is defined in a way that it rotates the hexagon by an angle of $120^{\circ}$. The corresponding rotational eigenvalues are $1$, $\omega$, $\omega^{2}$, where $\omega^3=1$. The chirality quantum number $\nu=0,1,-1$ correspond to the rotational eigenvalues $1$, $\omega$, $\omega^{2}$ respectively.

The lowest energy state is a singlet with $\nu=0$. The next higher energy level is sixfold degenerate, and it consists of two triplets, represented as $|\that_{m,\nu}\rangle$, given by $m=1,0,\onebar$ and $\nu=1,\onebar$. We neglect all the other higher energy states in a low energy theory. However, there are 3 more triplets (with $\nu=0$) which at $J_{F}\rightarrow\infty$ becomes degenerate with the 6 chiral triplets discusses here. But for any finite value of $J_F$ the gap between the two remains finite, hence not included in the present calculation. For these lowest energy singlet and triplet block eigenstates, we employ the same strategy as in the previous two subsections, and define a singlet creation operator $\hat{s}^{\dagger}$ and six triplet operators $t_{m,\nu}^{\dagger}$ in the Fock space  
with a constraint, 
$ \hat{s}^{\dagger}\hat{s}+\sum_{m,\nu}t_{m,\nu}^{\dagger}t_{m,\nu}=1. $
In terms of these singlet and triplet operators, the block Hamiltonian of up hexagon reads as: 
\begin{equation} \label{eq:uphexTS}
  \mathcal{H}_{\ahex}\approx E_{s}\shat^{\dagger}\shat+E_{t}\sum_{m}\sum_{\nu=\pm 1}\that_{m,\nu}^{\dagger}\that_{m,\nu},
\end{equation}
where $E_{s}$  and $ E_{t}$ are respectively the lowest singlet and triplet eigen-energies of the block Hamiltonian. 

With the block operators defined, we can now write the spin-$1/2$ operators on up hexagons as follows.
\begin{subequations} \label{eq:spinTS}
 \begin{align}
  S_{j}^{\alpha}&=-\sqrt{2}\sbar\left(\Re\left[\MC_{j}^{\alpha}\right]\Qhat_{\alpha 1}+\Im\left[\MC_{j}^{\alpha}\right]\Qhat_{\alpha \onebar}\right)~\mbox{for}~\alpha=x,y \\
S_{j}^{z}&=2\sbar\left(\Re\left[\MC_{j}^{z}\right]\Qhat_{z 1}-\Im\left[\MC_{j}^{z}\right]\Qhat_{z\onebar}\right)
 \end{align}
\end{subequations}
Here, $\Re\left[\MC\right]$ and $\Im\left[\MC\right]$ are the real and imaginary parts of of $\MC$. The $\MC_{j}^{\alpha}$ and $\MC_{j}^{z}$, which depend on $|J_{F}|/J_A$, are
\begin{subequations}
\begin{eqnarray}
 \MC_{j}^{\alpha}&=&\langle s|S_{j}^{+}|t_{11}\rangle=\langle s|S_{j}^{+}|t_{\onebar\onebar}\rangle ~\mbox{and} \\
 \MC_{j}^{z}&=&\langle s|S_{j}^{z}|t_{01}\rangle=\langle s|S_{j}^{z}|t_{0\onebar}\rangle.
\end{eqnarray}
\end{subequations}
Moreover, 
 $Q_{\alpha\nu} = \frac{1}{\sqrt{2}}\left(\that_{\alpha\nu}^{\dagger}+\that_{\alpha\nu}\right)$,
 $ P_{\alpha\nu} = \frac{i}{\sqrt{2}}\left(\that_{\alpha\nu}^{\dagger}-\that_{\alpha\nu}\right) $,
$\that_{x\nu} = \frac{1}{\sqrt{2}}\left(\that_{\onebar\nu}-\that_{1\nu}\right)$, 
 $\that_{y\nu} = \frac{i}{\sqrt{2}}\left(\that_{\onebar\nu}+\that_{1\nu}\right)$ and $ \that_{z\nu} = \that_{0\nu}$~\footnote{The  $\that_{m\nu}$ operators used to define $t_{\alpha\nu}$ are obtained via a simple rotation performed on the old $\that_{m\nu}$ operators. This transformation goes as: $\frac{1}{\sqrt{2}}\left(\that_{m\onebar}+\that_{m1}\right)  \rightarrow  \that_{m1}$ and $ \frac{1}{\sqrt{2}}\left(\that_{m\onebar}-\that_{m1}\right)  \rightarrow  \that_{m\onebar}$.}. This representation is very similar to the one derived recently in Ref.~\cite{Ghosh2016}. For $J_{F}\rightarrow -\infty$, it exactly becomes what is given in Ref.~\cite{Ghosh2016}. Moreover, the singlet operator on up-oriented hexagons is approximated by a mean-field $\bar{s}$, which describes the mean-field TS state.

Now in the full Hida model, we write all up hexagons as in Eq.~\ref{eq:uphexTS}, and the AFM bonds in the down hexagons using the spin representation of Eq.~\ref{eq:spinTS}. This turns the Hida model into the following triplon model:  
\begin{eqnarray}{\label{eq:H-trip}}
 \hat{H}_{t}^{TS}= e_{0}N_{uc} &+& \frac{1}{2}\sum_{\k}\sum_{\alpha=x,y,z}  
 \left[ \lambda\, {\bf \Phat}^{\dagger}_{\alpha}(\k) {\bf \Phat}_{\alpha}(\k)\right.\nonumber\\
 &+&\left. {\bf \Qhat}^{\dagger}_{\alpha}(\k)\,\mathcal{V}_{\alpha,\k}\,{\bf \Qhat}_{\alpha}(\k) \right].
\end{eqnarray}
Here, $N_{uc}$ is the total number of hexagonal unit-cells in the lattice, $e_{0}=\left(E_{s}-E_{t}\right)\bar{s}^{2}+\lambda\bar{s}^{2}+E_{t}-4\lambda$, and $\lambda$ is the Lagrange multiplier that is introduced to satisfy the local constraint, $\sbar^{2} + \sum_{\alpha\nu}\that^{\dagger}_{\alpha\nu}\that_{\alpha\nu} =1 $, on average. Moreover, 
 \begin{equation}
 {\bf \Qhat}_\alpha(\k) = \left[\begin{array}{c} \Qhat_{\alpha1}(\k) \\ \Qhat_{\alpha\onebar}(\k) \end{array}\right]~~\mbox{and}~~{\bf \Phat}_\alpha(\k) = \left[\begin{array}{c} \Phat_{\alpha1}(\k) \\ \Phat_{\alpha\onebar}(\k) \end{array}\right],
 \end{equation}
where $\Qhat_{\alpha 1}(\k)$ and $\Qhat_{\alpha\onebar}(\k)$ are the Fourier components of $\Qhat_{\alpha\nu}(\r)$. That is, $\Qhat_{\alpha\nu}(\r) = \frac{1}{\sqrt{N_{uc}}}\sum_\k e^{i\k\cdot\r}\Qhat_{\alpha\nu}(\k)$ for $\nu=1,\onebar$. Here, $\r$ denotes the position vector of the hexagonal units of the lattice, and $\k$ lies in the Brillouin zone of the corresponding reciprocal lattice. Likewise, $\Phat_{\alpha\nu}(\r) = \frac{1}{\sqrt{N_{uc}}}\sum_\k e^{i\k\cdot\r}\Phat_{\alpha\nu}(\k)$. Since $\Qhat_{\alpha\nu}(\r)$ and $\Phat_{\alpha\nu}(\r)$ are Hermitian, therefore, $\Qhat^{\dagger}_{\alpha\nu}(\k) = \Qhat_{\alpha\nu}(-\k)$ and $\Phat^{\dagger}_{\alpha\nu}(\k) = \Phat_{\alpha\nu}(-\k)$. Moreover, $[\Qhat_{\alpha\nu}(\k),\Phat_{\alpha^\prime \nu^\prime}(\k^\prime)] = i\delta_{\alpha\alpha^\prime}\delta_{\nu\nu^\prime}\delta_{\k+\k^\prime=0}$, while the $\Qhat_{\alpha\nu}(\k)$'s commute amongst themselves and the same for $\Phat_{\alpha\nu}(\k)$'s.

The $\hat{H}^{TS}_t$ is a problem of two oscillators for each $\alpha$ described by $\Qhat_{\alpha1}(\k)$ and $\Qhat_{\alpha\onebar}(\k)$, and coupled via 
\begin{equation}
\mathcal{V}_{\alpha,\k}=\left[\begin{array}{lcl}  \lambda-2\sbar^2\epsilon_{\alpha1,\k} && \sbar^2\eta_{\alpha,\k} \\ 
\sbar^2\eta^*_{\alpha,\k} && \lambda-2\sbar^2\epsilon_{\alpha\onebar,\k} \end{array}\right].
\label{eq:V}
\end{equation}
The $\mathcal{V}_{\alpha,\k}$ is a Hermitian matrix, with $\eta^*_{\alpha,\k}$ as the complex conjugate of $\eta_{\alpha,\k}$. The $\epsilon_{\alpha\nu,\k}$ and $\eta_{\alpha,\k}$ are given below.
\begin{widetext}
\begin{subequations}
\begin{align}
 \epsilon_{\alpha 1,\k} =&~  2J_{A}\big(\Re\left[\MC_{3}^{\alpha}\right]\Re\left[\MC_{6}^{\alpha}\right]\cos\k\cdot\a_2+\Re\left[\MC_{1}^{\alpha}\right]\Re\left[\MC_{4}^{\alpha}\right]\cos{\k\cdot\a_1}+\Re\left[\MC_{2}^{\alpha}\right]\Re\left[\MC_{5}^{\alpha}\right]\cos\k\cdot\a_{3}\big) \\
\epsilon_{\alpha\onebar,\k} =&~  2J_{A}\big(\Im\left[\MC_{3}^{\alpha}\right]\Im\left[\MC_{6}^{\alpha}\right]\cos\k\cdot\a_2+\Im\left[\MC_{1}^{\alpha}\right]\Im\left[\MC_{4}^{\alpha}\right]\cos\k\cdot\a_1+\Im\left[\MC_{2}^{\alpha}\right]\Im\left[\MC_{5}^{\alpha}\right]\cos\k\cdot\a_{3}\big) \\
\eta_{\alpha,\k} =&~  J_{A}\big(\Re\left[\MC_{6}^{\alpha}\right]\Im\left[\MC_{3}^{\alpha}\right]e^{i\k\cdot\bf{a_{1}}}+\Im\left[\MC_{6}^{\alpha}\right]\Re\left[\MC_{3}^{\alpha}\right]e^{-i\k\cdot\bf{a_{1}}}+\Re\left[\MC_{4}^{\alpha}\right]\Im\left[\MC_{1}^{\alpha}\right]e^{i\k\cdot\bf{a_{2}}}
+\Im\left[\MC_{4}^{\alpha}\right]\Re\left[\MC_{1}^{\alpha}\right]e^{-i\k\cdot\bf{a_{2}}}\nonumber \\
&~ +\Re\left[\MC_{5}^{\alpha}\right]\Im\left[\MC_{2}^{\alpha}\right]e^{i\k\cdot\a_{3}}+\Im\left[\MC_{5}^{\alpha}\right]\Re\left[\MC_{2}^{\alpha}\right]e^{-i\k\cdot\a_{3}}\big) 
\end{align}
\end{subequations}
\end{widetext}
As in Ref.~\cite{Ghosh2016}, the $\hat{H}^{TS}_t$ can be diagonalized by a unitary rotation of $\Qhat_{\alpha 1}(\k)$ and $\Qhat_{\alpha\onebar}(\k)$ to the new operators, $\Qhat_{\alpha +}(\k)$ and $\Qhat_{\alpha-}(\k)$.
 \begin{equation}
 \left[\begin{array}{c} \Qhat_{\alpha+}(\k) \\ \Qhat_{\alpha-}(\k) \end{array}\right] = \mathcal{U}^{\dagger}_{\alpha,\k} \left[\begin{array}{c} \Qhat_{\alpha1}(\k) \\ \Qhat_{\alpha\onebar}(\k) \end{array}\right] \label{eq:UQ}
 \end{equation}
The unitary matrix $\mathcal{U}_{\alpha,\k}$ is given as:
\begin{equation}
\mathcal{U}_{\alpha,\k}=\left[\begin{array}{lcr} \cos{\frac{\theta_{\alpha,\k}}{2}} && - e^{-i\phi_{\alpha,\k}}\sin{\frac{\theta_{\alpha,\k}}{2}} \\ 
 e^{i\phi_{\alpha,\k}}\sin{\frac{\theta_{\alpha,\k}}{2}} && \cos{\frac{\theta_{\alpha,\k}}{2}} \end{array}\right],
\end{equation}
where $\theta_{\alpha,\k} = \tan^{-1}{\{|\eta_{\alpha,\k}|/(\epsilon_{\alpha\onebar,\k}-\epsilon_{\alpha 1,\k})\}}$, and $\eta_{\alpha,\k} = |\eta_{\alpha,\k}| e^{-i\phi_{\alpha,\k}}$ with $|\eta_{\alpha,-\k}|=|\eta_{\alpha,\k}|$ and $\phi_{\alpha,-\k} = -\phi_{\alpha,\k}$. 

The $\hat{H}^{TS}_t$ in the diagonal form can be written as:
\begin{equation}\label{eq:Ht-diagonal}
  \hat{H}^{TS}_t=e_{0}N_{uc} + \sum_{\k}\sum_{\alpha=x,y,z}\sum_{\mu=\pm}E_{\alpha\mu,\k}\left[ \that^{\dagger}_{\alpha\mu}(\k) \that_{\alpha\mu}(\k)+\frac{1}{2}  \right],
\end{equation}
where $ \that_{\alpha\mu}(\k) = \sqrt{\frac{E_{\alpha\mu,\k}}{2\lambda}}\Qhat_{\alpha\mu}(\k) + i\sqrt{\frac{\lambda}{2E_{\alpha\mu,\k}}}\Phat_{\alpha\mu}(\k) $ are the renormalized triplon operators, and 
$E_{\alpha\mu,\k}=\sqrt{\lambda(\lambda-2\sbar^2\xi_{\alpha\mu,\k})}$ are the triplon energy dispersions with $ \xi_{\alpha\mu,\k} = [(\epsilon_{\alpha\onebar,\k} + \epsilon_{\alpha1,\k})-\mu\sqrt{(\epsilon_{\alpha\onebar,\k}-\epsilon_{\alpha1,\k})^2+|\eta_{\alpha,\k}|^2}]/2$. The label, $\mu=\pm$, for new operators  defined in Eqs.~(\ref{eq:UQ}), is analogous to, but different from, the old $\nu$. The ground state energy per unit-cell from Eq.~(\ref{eq:Ht-diagonal}) is given by
\begin{equation}
 e^{TS}_g = e_{0}+\frac{1}{2N_{uc}}\sum_\k\sum_{\alpha=x,y,z}\sum_{\mu=\pm} E_{\alpha\mu,\k}.\label{eq:eg}
\end{equation}  
Again, by minimizing $e^{TS}_g$ with respect to $\lambda$ and $\sbar^2$, we get the following mean-field equations.
\begin{subequations}  \label{sc-TS}
 \begin{align}
  \sbar^2 &= 4-\frac{1}{2N_{uc}}\sum_{\k}\sum_{\alpha=x,y,z}\sum_{\mu=\pm}\frac{\lambda-\sbar^{2}\xi_{\alpha\mu,\k}}{E_{\alpha\mu,\k}} \label{eq:sbar} \\
   \lambda &=\left(E_{t}-E_{s}\right)+\frac{\lambda}{2N_{uc}}\sum_{\k}\sum_{\alpha=x,y,z}\sum_{\mu=\pm}\frac{\xi_{\alpha\mu,\k}}{E_{\alpha\mu,\k}} \label{eq:lam}
 \end{align}
\end{subequations}

Having thus formulated the TMFT's for the Hida model with respect to the physically motivated HS, D-HS and TS states, we next discuss the results of these calculations, in particular, the competition between the three candidate states to be the ground state.

\subsection{Results from Triplon Mean-Field Calculations}
\begin{figure}[t]
\includegraphics[width=0.9\columnwidth]{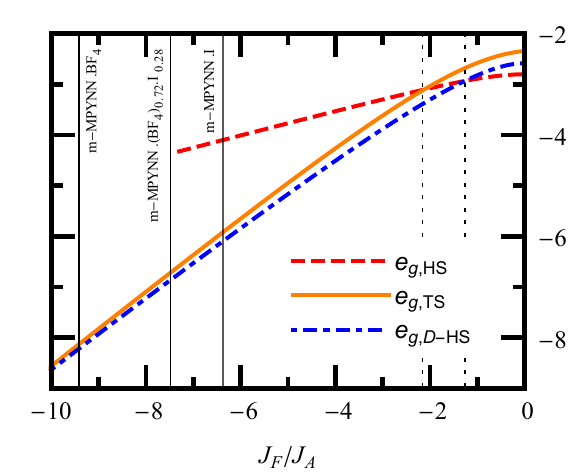}
\caption{\label{hidaE}
The triplon mean-field energies of the HS (hexagonal singlet), D-HS (dimerized HS) and TS (trimerized singlet) states of the Hida model.} 
\end{figure}
\begin{figure*}[t]
\includegraphics[width=\textwidth]{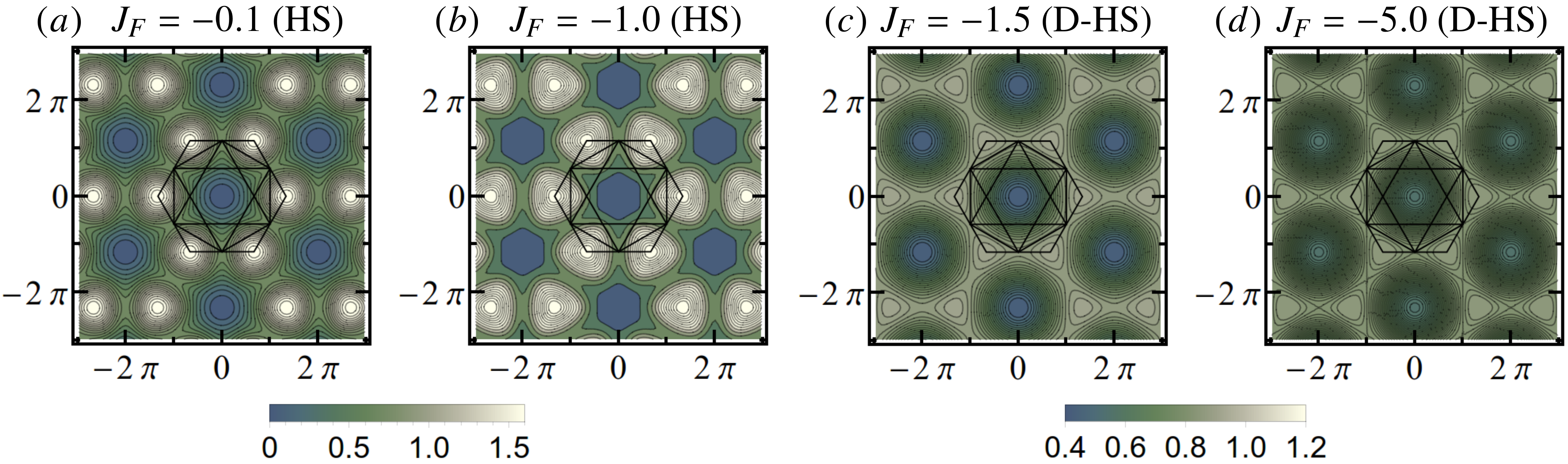}
\caption{\label{Sq_HS_DHS}
The static structure factor, $S(\q)$, in the HS and D-HS ground states of the Hida model. The color-code bar on the bottom-left applies to the plots ($a$) and ($b$), and that on the bottom-right applies to ($c$) and ($d$). The extended Brillouin zone (upto 4th Brillouin zone) is shown in solid-black lines.} 
\end{figure*}

We self-consistently solve Eqs.~\eqref{sc-HSS},~\eqref{sc-DHSS} and~\eqref{sc-TS} for the three cases as a function of the negative $J_F$, with $J_A=1$ in the calculations. This allows us to compute the energies, $e_{g}^{HS}$, $e_{g}^{D\mbox{-}HS}$ and $e_{g}^{TS}$, of the three states. A comparison of these energies would give us some understanding of the possible phase transitions with change in $J_F$ in the ground state of the Hida model. 

The competition between one symmetry preserving (HS) and two symmetry breaking (D-HS and TS) phases is shown in Fig.~\ref{hidaE}. For small $J_F$, the HS state is expectedly the ground state of the system. However as $|J_F|/J_A$ increases, interestingly the symmetry-breaking (D-HS and TS) phases become lower in energy. In particular, it is the D-HS state which first becomes lower in energy than the HS state at $J_{F} \approx -1.26$. While the TS state also crosses the HS state at $J_F \approx -2.15$, but it never crosses the D-HS state. In fact, for large negative $J_F$, the TS asymptotically approaches the D-HS state. That is, the TS and D-HS represent the same physical state for large negative $J_F$. Thus, according to the triplon mean-field theory, the D-HS is the ground state of the Hida model for $|J_F|/J_A  > 1.26$. For large negative $J_F$, the D-HS is same as the TS state. It is consistent with the fact that for spin-1 KHA, which is the large negative $J_F$ limit of the Hida model, the TS is the ground state. 

To see the implications of our finding for real materials, we also indicate the positions of different \ce{m-MPYNN.X} salts (which motivated the Hida model in the first place) on $J_F/J_A$ axis in Fig.~\ref{hidaE}. Our triplon analysis clearly suggests that these organic salts at low temperatures would realize the symmetry-breaking D-HS phase, as opposed to the uniform HS phase proposed by Hida in his original paper. Notably, consistent with the known behavior of this family of materials, the D-HS ground state also has a finite spin-gap for the entire range of $J_F$.

\begin{figure}[t]
\includegraphics[width=0.9\columnwidth]{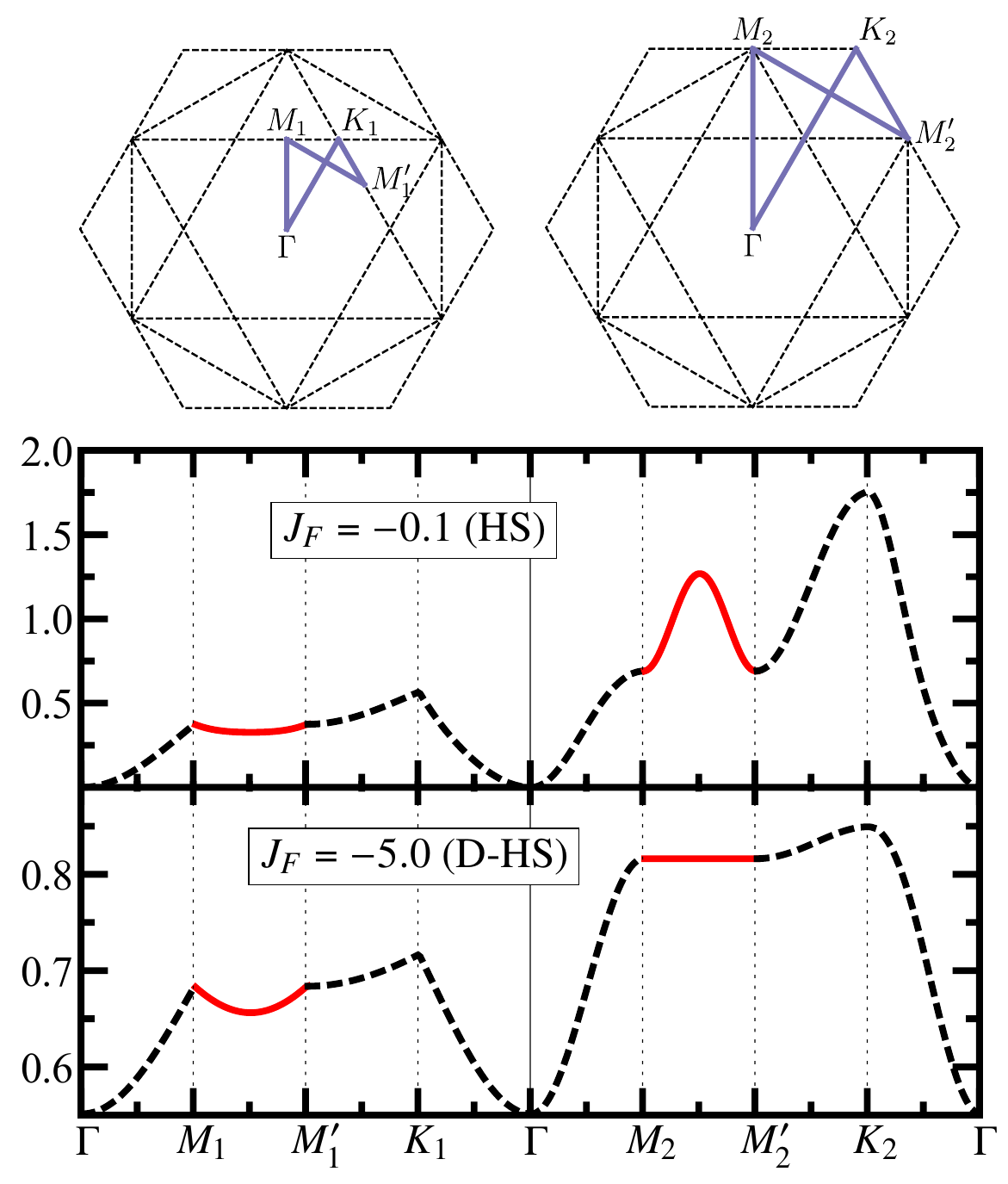}
\caption{\label{Sq_HS_DHS_cut}
The static structure factor, $S(\q)$, plotted along the high-symmetry lines of the first and the extended Brillouin zone in the HS and D-HS ground states of the Hida model. The high-symmetry lines are shown in blue over the Brillouin zones above the actual plots.} 
\end{figure}

Since the HS, D-HS and TS states are all spin-gapped and non-magnetic, usual thermodynamics measurements can not distinguish between them. But neutron diffraction may tell us more precisely as to which of these is the low temperature phase of the \ce{m-MPYNN.X} salts. To this end, we calculate the static structure factor, $S(\q)$, of these three states for different values of $J_F$. It is defined as: $ S\left(\q\right)=\frac{1}{N}\sum_{i,j} \langle \vec{S}_{i}\cdot\vec{S}_{j} \rangle e^{-i\q.\bf{r}_{ij}}$, where $N$ is the total number of lattice sites, and $\vec{S}_{i}$ and $\vec{S}_{j}$ are two spins on the lattice sites separated by a distance $\bf{r}_{ij}$. The $i,j$ sum here runs over all the lattice sites. 

\begin{figure*}
\includegraphics[width=.95\textwidth]{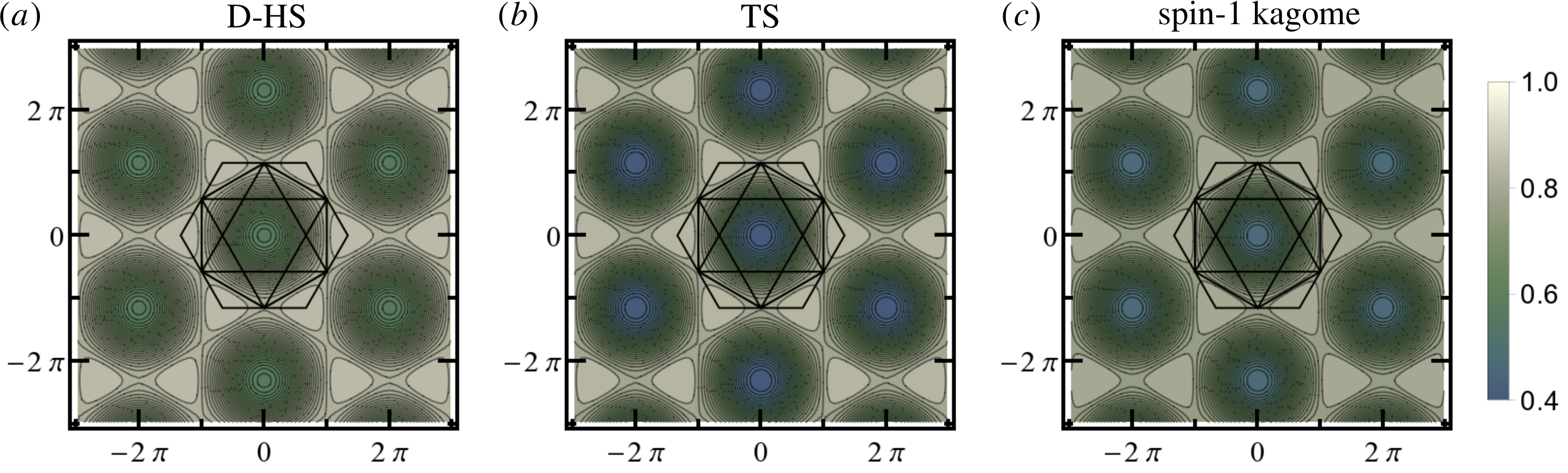}
\caption{\label{Sq_DHS_TS_Kagome}
Comparing the $S(\q)$ in the D-HS and TS states of the Hida model (at $J_{F}/J_{A}=-10$) with that in the TS ground state of the spin-1 kagome Heisenberg antiferromagnet.} 
\end{figure*}

In Fig.~\ref{Sq_HS_DHS}, we present the intensity contourplots of $S(\q)$ for such values of $J_F$ where either HS or D-HS forms the ground state of Hida model within TMFT (see Fig.~\ref{hidaE}). In all of these plots, the intensity maxima always occur at the corners of the fourth Brillouin Zone (BZ4) with $|\q|=4\pi/3$, while the intensity is minimum at the zone center. But there are some notable features that can visibly distinguish between HS and D-HS. 
An important distinction between the two states comes from the curvature of the intensity contours around minima (zone center). For $S(\q)$ in the HS state, the contours enclosing the zone center become negatively curved (concave) as one moves away from the center, while they are always positively curved (convex) in the D-HS state. This distinction is further linked to the shape of intensity contours around the points of maxima (the corners of BZ4), which are curved triangles pointing away from the points of minima in the D-HS state, and pointing towards the points of minima in the HS state. These curved triangular contours form a kagom\'e like pattern in the $S(\q)$ of D-HS state but not in the HS state, while the points of maxima in both form a honeycomb lattice. 

To make the distinguishing feature more precise, in Fig. \ref{Sq_HS_DHS_cut}, we plot the $S(\q)$ in the HS and D-HS phases along certain high-symmetry directions in the first and the extended Brillouin zones. The qualitative difference in $S(\q)$ between the two phases shows up along the $M_{1}M_{1}^{\prime}$ line in the first Brillouin zone and $M_{2}M_{2}^{\prime}$ line in the extended Brillouin zone. In the HS phase, the variation of $S(\q)$ along $M_{1}M_{1}^{\prime}$ is very small (mostly flat except near the two ends), while its variation along $M_{2}M_{2}^{\prime}$ is significant. Interestingly, it is exactly opposite in the D-HS phase, where the $S(\q)$ along $M_{2}M_{2}^{\prime}$ stays pretty flat, while it varies significantly along the $M_{1}M_{1}^{\prime}$ line. To quantify this relative variation, we define a quantity  
\begin{equation*}
f_{v}=\frac{\max[S(\q)_{M_{1}M_{1}^{\prime}}]-\min[S(\q)_{M_{1}M_{1}^{\prime}}]}{\max[S(\q)_{M_{2}M_{2}^{\prime}}]-\min[S(\q)_{M_{2}M_{2}^{\prime}}]} 
\end{equation*}
as the ratio of the difference between the maximum and minumum values of $S(\q)$ along $M_1M_1^\prime$ and $M_2M_2^\prime$ directions. We for instance get $f_{v}\sim 0.0789$ in the HS phase for $J_{F}=-0.1$, and $f_{v}\sim 8471.62$ for $J_{F}=-5.0$ in the D-HS phase. Clearly, for experiments, it suggests that if $f_v <1$, then the material is in the HS phase, and if $f_v >1$, then it's in the D-HS phase. We propose this relative variation of $S(\q)$ along $M_{1}M_{1}^{\prime}$ and $M_{2}M_{2}^{\prime}$ directions as a characteristic feature that can unambiguously differentiate between the HS and D-HS states in a neutron diffraction experiment.

Since the TS state for large negative $J_F$ approaches the D-HS state, in Fig.~\ref{Sq_DHS_TS_Kagome} we also compare the $S(\q)$ in the D-HS state with that in the TS state at $J_{F}/J_{A}=-10$, and also with the $S(\q)$ in the TS ground state of the spin-$1$ kagom\'e Heisenberg AFM model (studied in our earlier paper~\cite{Ghosh2016}). Well, they all look pretty much the same! This clearly implies that the TS ground state that we and others have found for the spin-1 KHA is essentially the D-HS ground state of the Hida model in the limit of large negative $J_F$.  Or in other words, the TS ground state of the spin-1 KHA is adiabatically connected to the D-HS ground state of the spin-1/2 Hida model.

Clearly, the TMFT of Hida model has given us an important understanding of the ground state. But, as is the case with triplon analysis, it (like the spin-wave analysis) is based on an a-priori knowledge or insights about the possible ground state. That is why, we first motivated the three states (HS, D-HS and TS), and then studied their competition for the ground state by doing triplon analysis. However, it would be nice, if we could also arrive at the same or similar physical conclusions by some alternate method without much a-priori assumptions about the possible ground state. With this motivation, we further investigate the Hida model by doing Schwinger boson mean-field theory (SBMFT) in the next section. Interestingly, the results of SBMFT qualitatively agree with what TMFT has taught us, and also reveal some novel features of the HS (for small $J_F$) to TS (for large $J_F$) transition. 

\section{Schwinger Boson Mean-Field Theory}\label{sec:SB_method}
The SBMFT (Schwinger boson mean-field theory) has been proven to be effective in describing the ordered and disordered phases of interacting quantum spins~\cite{Arovas1988, Auerbach1988, HRK-SBMFT, Auerbach-book}. It has been applied to the Heisenberg models on different lattices such as the square \cite{Wei1994}, triangular \cite{Sachdev1992, Wang2006} and kagom\'e lattices~\cite{Sachdev1992, Manuel1994, Wang2006, Li2007, Messio2010, Mondal2017}.     
Here, we formulate the SBMFT for the Hida model of Eq.~\eqref{eq:Hamil}. 
We start by writing the Schwinger boson representation of the spin operators in terms of the boson operators, $a_{i}$ and $b_{i}$ defined on every site, $i$, as:
\begin{subequations}
\begin{align}\label{orders}
& S^{+}_{i} = a^{\dagger}_{i}b_{i}\\
& S^{-}_{i} = b^{\dagger}_{i}a_{i}\\
& S^{z}_{i} = \frac{1}{2}\left(a^{\dagger}_{i}a_{i}-b^{\dagger}_{i}b_{i}\right)
\end{align}
\end{subequations}
with the constraint,
$ \aidag a_{i}+\bidag b_{i}=2S $,
for the spin quantum number, $S$. For the moment, we keep $S$ as general, but eventually, we will consider $S=1/2$. 

To write the Heisenberg exchange interaction between the spins, $\vec{S}_i$ and $\vec{S}_j$, 
we introduce the following two bosonic operators involving the sites $i$ and $j$. 
\begin{subequations}
\begin{align}
& A_{ij}=\frac{1}{\sqrt{2}}\left(a_{i}b_{j}-b_{i}a_{j}\right) \\
& F_{ij}=\frac{1}{\sqrt{2}}\left(\aidag a_{j} +\bidag b_{j} \right) 
\end{align}
\end{subequations}
Physically, the $F_{ij}$ represents the hopping of the bosons and $A_{ij}^{\dagger}$ forms the singlet between $i$-th and $j$-th sites. Using these two operators, the Heisenberg exchange operator can be written as, $\vec{S}_i\cdot\vec{S}_j = \, :F^{\dagger}_{ij}F_{ij}: -A^{\dagger}_{ij}A_{ij} $, where $:\mathcal{O}:$ is the normal ordered form of the operator $\mathcal{O}$. Given this operator identity, 
the Hida model of Eq.~\eqref{eq:Hamil} takes the following form. 
\begin{eqnarray}\label{eq:H-SB}
H = &&\frac{J_{F}}{2} \sum_{{\langle i,j \rangle}}^{\fer} \left(:F^{\dagger}_{ij}F_{ij}:-A^{\dagger}_{ij}A_{ij}\right)\nonumber\\
&+&\frac{J_{A}}{2} \sum_{{\langle i,j \rangle}}^{\afer} \left(:F^{\dagger}_{ij}F_{ij}:-A^{\dagger}_{ij}A_{ij}\right)
\end{eqnarray}
Due to the local constraint on the number of bosons per site, the $A_{ij}$ and $F_{ij}$ are also constrained to satisfy the condition
\begin{equation}\label{cons2}
:F^{\dagger}_{ij}F_{ij}:+A^{\dagger}_{ij}A_{ij}=2S^{2}.
\end{equation}

Since Eq.~\eqref{eq:H-SB} is quartic in Schwinger bosons, we decouple the operators there  by introducing the mean fields, $\alpha_F$, $\phi_F$, $\alpha$, $\phi$, $\alpha^{\prime}$ and $\phi^{\prime}$, which are defined as follows: on all FM bonds ($\fer$), $\alpha_{F}=\langle A_{ij} \rangle$ and $\phi_{F}=\langle F_{ij} \rangle$, on every AFM bond of the up-oriented hexagons ($\afup$), $\alpha_{A}=\langle A_{ij} \rangle$ and $\phi_{A}=\langle F_{ij} \rangle$, and on every AFM bond of the down-oriented hexagons ($\afdn$), $\alpha^{\prime}_{A}=\langle A_{ij} \rangle$ and $\phi^{\prime}_{A}=\langle F_{ij} \rangle$. This choice of mean-field parameters is the very minimal that would allow spontaneous symmetry-breaking (dimerization of the AFM hexagons), not by a-priori assumption, but by the self-consistent determination of $(\alpha_{A}, \phi_{A})$ and $(\alpha^{\prime}_{A}, \phi^{\prime}_A)$ through the mean-field dynamics of the Schwinger bosons. So, if it turns out that $(\alpha_{A}, \phi_{A})=(\alpha^{\prime}_{A},\phi^{\prime}_{A})$, then we have the uniform HS phase. But if $(\alpha_{A}, \phi_{A}) \ne (\alpha^{\prime}_{A},\phi^{\prime}_{A})$, then we have the D-HS phase. For simplicity, we treat these mean-field parameters as real. 

Under this mean-field approximation, the Hamiltonian in Eq.~\eqref{eq:H-SB} takes the following form.
\begin{eqnarray}\label{eq:Hamil-SBMFT}
 \mathcal{H}_{MF}^{SB}=&&\frac{J_F}{2}\sum_{{\langle i,j \rangle}}^{\fer}\left[\phi_{F}\left(F^{\dagger}_{ij}+F_{ij}\right)-\alpha_{F}\left(A_{ij}^{\dagger}+A_{ij}\right)\right]\nonumber\\
 &+&\frac{J_A}{2}\sum_{{\langle i,j \rangle}}^{\afup}\left[\phi_{A}\left(F^{\dagger}_{ij}+F_{ij}\right)-\alpha_{A}\left(A_{ij}^{\dagger}+A_{ij}\right)\right]\nonumber\\
 &+&\frac{J_A}{2}\sum_{{\langle i,j \rangle}}^{\afdn}\left[\phi^{\prime}_{A}\left(F^{\dagger}_{ij}+F_{ij}\right)-\alpha^{\prime}_{A}\left(A_{ij}^{\dagger}+A_{ij}\right)\right]\nonumber\\
 &-&\frac{3J_F}{2}N_{uc}\left(\phi_{F}^{2}-\alpha_{F}^{2}\right)\nonumber\\
 &-&\frac{3J_{A}}{2}N_{uc}\left[\left(\phi_{A}^{2}-\alpha_{A}^{2}\right)+\left({\phi^{\prime}_{A}}^{2}-{\alpha^{\prime}_{A}}^{2}\right)\right]\nonumber\\
 &+&\lambda\sum_{i}\left(\aidag a_{i}+\bidag b_{i}-1\right)
\end{eqnarray}
Here, the last term imposes the local number constraint, $\aidag a_{i}+\bidag b_{i}=2S$ for $S=1/2$, only on average through Lagrange multiplier, $\lambda$.

Although bilinear in Schwinger boson operators, the mean-field problem in Eq.~\eqref{eq:Hamil-SBMFT} needs to be handled carefully. To proceed, we first consider an isolated FM bond, say, the $1$-$2$ bond in Fig.~\ref{hidalattice}. The Hamiltonian of this FM bond in SBMFT reads as:
\begin{eqnarray}
 \mathcal{H}_{MF,12}^{SB}&=&\frac{J_F}{2}\left[\phi_{F}\left(F_{12}^{\dagger}+F_{12}\right)-\alpha_{F}\left(A_{12}^{\dagger}+A_{12}\right)\right.\nonumber\\
 &-&\left.\phi_F^{2}+\alpha_F^{2}\right]+\lambda\sum_{i=1,2}\left(\aidag a_{i}+\bidag b_{i}-1\right).
\end{eqnarray}
We diagonalize it by applying the Bogoliubov transformation, 
\[ \left[ \begin{array}{c}
a_{1} \\
a_{2}\\
b_{1}^{\dagger} \\
b_{2}^{\dagger}\end{array} \right] = \mathcal{U}\left[ 
\begin{array}{c}
\tilde{a}_{-} \\
\tilde{a}_{+}\\
\tilde{b}_{-}^{\dagger} \\
\tilde{b}_{+}^{\dagger}\end{array} 
 \right]. \]
where,
\[ \mathcal{U} = \frac{1}{\sqrt{2}}\left[ \begin{array}{cccc}
\cosh\eta & \cosh\eta & -\sinh\eta & \sinh\eta \\
\cosh\eta & -\cosh\eta & \sinh\eta & \sinh\eta \\
\sinh\eta & -\sinh\eta & \cosh\eta & \cosh\eta\\
-\sinh\eta & -\sinh\eta & \cosh\eta & -\cosh\eta\end{array}\right]\]
for $\tanh2\eta=-J_{F}\alpha_{F}/2\sqrt{2}\lambda$. Under this transformation, the FM bond Hamiltonian reads as:
\begin{eqnarray}
 \mathcal{H}_{MF,12}^{SB}&=& \omega_{-}\left(\tilde{a}_{-}^{\dagger}\tilde{a}_{-}+\tilde{b}_{-}^{\dagger}\tilde{b}_{-}\right)+\omega_{+}\left(\tilde{a}_{+}^{\dagger}\tilde{a}_{+}+\tilde{b}_{+}^{\dagger}\tilde{b}_{+}\right)  \nonumber\\
 &\ &-\frac{J_F}{2}\left(\phi_F^{2}-\alpha_F^{2}\right)+(\omega_{+}+\omega_{-})-4\lambda \label{eq:FMbond}
 \end{eqnarray}
where 
\begin{equation}
 \omega_{\pm}=\pm \frac{|J_{F}|\phi_{F}}{2\sqrt{2}}+\sqrt{\lambda^{2}-\left(\frac{J_{F}\alpha_{F}}{2\sqrt{2}}\right)^{2}}.
\end{equation}
Importantly, we recognize that the operators, $\tilde{a}_{-}$ and $\tilde{b}_{-}$, with energy $\omega_{-}$, will have to condense in order to form a bound moment (spin-1) in the ground state. Therefore, we treat the operators $\tilde{a}_{-}$ and $\tilde{b}_{-}$ as average amplitudes $\bar{a}$ and $\bar{b}$, respectively. 
We do the same treatment of all the FM bonds of the Hida model, taking the same average amplitudes, $\bar{a}$ and $\bar{b}$. Moreover, we drop the subscript, $+$, from  $\tilde{a}_{+}$ and $\tilde{b}_{+}$ (which is not essential anymore), and label these operators by the bond index, \rom{1}, \rom{2} and \rom{3} for the FM bonds 1-2, 3-4 and 5-6, respectively.  

As a result of the above treatment, the full Hamiltonian in Eq.~\ref{eq:Hamil-SBMFT} will also acquire linear terms in $\tilde{a}_i$'s and $\tilde{b}_i$'s (for $i = \rom{1}$, $\rom{2}$ and $\rom{3}$), in addition to the bilinear terms. We get rid of these linear terms by making the following displacement transformation: 
\[ \left[ \begin{array}{c}
\widetilde{a}_{i} \\
\widetilde{b}_{i}\end{array} \right] \rightarrow \left[ \begin{array}{c}
\tilde{a}_{i}  + r \bar{b} \\
\tilde{b}_{i}  + r \bar{a} \end{array} \right],\] 
where
$
r =-\frac{J_{A}}{2\sqrt{2}} (\alpha_{A}+\alpha^{\prime}_{A})/[\omega_{+}-\frac{J_{A}}{2\sqrt{2}}(\phi_{A}+\phi^{\prime}_{A})] $.
Next we do the Fourier transformation as follows:
\begin{subequations}
 \begin{eqnarray}
  \tilde{a}_{i} (\r) &=& \frac{1}{\sqrt{N_{uc}}}\sum_{\k}  \tilde{a}_{i,\k}  e^{-i\k\cdot\r},\\
   \tilde{b}_{i} (\r) &=& \frac{1}{\sqrt{N_{uc}}}\sum_{\k}  \tilde{b}_{i,\k} e^{i\k\cdot\r}.
 \end{eqnarray}
 \end{subequations}
The resulting SBMFT Hamiltonian in the momentum space can be written as:
\begin{equation}\label{eq:hamilmf}
\mathcal{H}_{MF}^{SB}=\sum_{\k}\left[ \begin{array}{cc}
\widetilde{\bf{a}}_{\k}^{\dagger} & \widetilde{\bf{b}}_{\k}\end{array}\right]\left[ \begin{array}{cc}
\mathcal{V}_{1\k} & \mathcal{V}_{2\k}   \\
\mathcal{V}_{2\k}^{\dagger} & \mathcal{V}_{1\k}  \end{array}\right]\left[ \begin{array}{c}
\widetilde{\bf{a}}_{\k}\\
\widetilde{\bf{b}}_{\k}^{\dagger}  \end{array}\right]+e_{0}N_{uc},
\end{equation}
where $\tilde{\bf{a}}_{\k}^{\dagger}=\left[\tilde{a}_{\rom{1},\k}^{\dagger}\ \tilde{a}_{\rom{2},\k}^{\dagger}\ \tilde{a}_{\rom{3},\k}^{\dagger}\right]$, $\tilde{\bf{b}}_{\k}=\left[\tilde{b}_{\rom{1},\k}\ \tilde{b}_{\rom{2},\k}\ \tilde{b}_{\rom{3},\k}\right]$,
\begin{equation}
\mathcal{V}_{1\k}=\left[ \begin{array}{ccc}
 \omega_{+} & \Delta+\Delta^{\prime} e^{-i\k\cdot\a_{2}} & \Delta+\Delta^{\prime} e^{-i\k\cdot\a_{3}}   \\
\Delta+\Delta^{\prime} e^{i\k\cdot\a_{2}} &  \omega_{+} & \Delta+\Delta^{\prime} e^{-i\k\cdot\a_{1}}\\
\Delta+\Delta^{\prime} e^{i\k\cdot\a_{3}}& \Delta+\Delta^{\prime} e^{i\k\cdot\a_{1}} & \omega_{+}\end{array}\right]
\end{equation}
and
\begin{equation}
\mathcal{V}_{2\k}^{\dagger}=\left[ \begin{array}{ccc}
0 & -t+t^{\prime}e^{i\k\cdot\a_{2}} & t-t^{\prime}e^{i\k\cdot\a_{3}}   \\
t-t^{\prime}e^{-i\k\cdot\a_{2}} &  0 & -t+t^{\prime}e^{i\k\cdot\a_{1}}\\
-t+t^{\prime}e^{-i\k\cdot\a_{3}} & t-t^{\prime}e^{-i\k\cdot\a_{1}} & 0\end{array}\right]
\end{equation}

\begin{eqnarray}
 t&=&\frac{J_{A}}{4\sqrt{2}}\left(-\alpha_{A}+\phi_{A}\sinh2\eta\right), \nonumber\\
 t^{\prime}&=&\frac{J_{A}}{4\sqrt{2}}\left(-\alpha^{\prime}_{A}+\phi^{\prime}_{A}\sinh2\eta\right), \nonumber\\
 \Delta&=&\frac{J_{A}}{4\sqrt{2}}\left(-\alpha_{A}\sinh2\eta-\phi_{A}\right) ~\mbox{and} \nonumber\\
 \Delta^{\prime}&=&\frac{J_{A}}{4\sqrt{2}}\left(-\alpha^{\prime}_{A}\sinh2\eta-\phi^{\prime}_{A}\right). \nonumber
\end{eqnarray}
Here, $\a_{1}$ and $\a_{2}$ are lattice vectors as given in Fig.~\ref{hidalattice}.
In Eq.~\ref{eq:hamilmf}, the constant, $e_0$, is given by
\begin{widetext}
 \begin{eqnarray}
 e_{0}=3&&\left[-\omega_{+}-2\lambda(2S+1)-\frac{J_F}{2}\left(\phi_{F}^{2}-\alpha_{F}^{2}\right)
 -\frac{J_{A}}{2}\left[\left(\phi_{A}^{2}-\alpha_{A}^{2}\right)+\left({\phi^{\prime}_{A}}^{2}-{\alpha^{\prime}_{A}}^{2}\right)\right]+2\sqrt{\lambda^{2}-\frac{1}{2}J_{F}^{2}\alpha_{F}^{2}}\right.\nonumber\\
 &&+\left\{\omega_{+}+\frac{J_A}{2\sqrt{2}}\left(\phi_{A}+\phi^{\prime}_{A}\right)-\frac{\left(\frac{J_A}{2\sqrt{2}}\left(\alpha_{A}+\alpha^{\prime}_{A}\right)\right)^{2}}{\omega_{+}-\frac{J_A}{2\sqrt{2}}\left(\phi_{A}-\phi^{\prime}_{A}\right)}\right\}\bar{\rho}^{2}\nonumber\\
&&\left.-\frac{J_A}{2\sqrt{2}}\left(\alpha_{A}+\alpha^{\prime}_{A}\right)\sinh2\eta \left\{1+\left(\frac{\frac{J_A}{2\sqrt{2}}\left(\alpha_{A}+\alpha^{\prime}_{A}\right)}{\omega_{+}-\frac{J_A}{2\sqrt{2}}\left(\phi_{A}-\phi^{\prime}_{A}\right)}\right)^{2}\right\}\bar{\rho}^{2}\right]. 
\end{eqnarray}
\end{widetext}
The $\bar{\rho}^{2} = \bar{a}^2+\bar{b}^2$ is a measure of the moment formation per FM bond. For large negative $J_F$, which corresponds to having spin-1 moment per FM bond, the $\bar{\rho}^2$ must tend to the value of $2$ (which it does in our calculations).

We diagonalize Eq.~\eqref{eq:hamilmf} using Bogoliubov transformation. In the diagonal form, it reads as: \[\mathcal{H}_{MF}^{SB} = \sum_{\k}\sum_{i=1}^{3}E_{i,\k}(\text{a}_{i,\k}^{\dagger}\text{a}_{i,\k}+\text{b}_{i,\k}\text{b}_{i,\k}^{\dagger})+e_{0}N_{uc},\] where $E_{i,\k}$'s are the six quasiparticle dispersions for six boson modes per unit-cell. Its ground state is the vacuum of the Bogoliubov quasiparticles, and the ground state energy per unit cell is given as:
\begin{equation}
 e_{g}=\frac{1}{2N_{uc}}\sum_{\k}\sum_{i=1}^{6}E_{i,\k}+e_{0}, 
\end{equation}
which is a function of the mean-field parameters, $\alpha_{F}$, $\phi_{F}$, $\alpha_{A}$, $\phi_{A}$, $\alpha^\prime_{A}$, $\phi^\prime_{A}$, the Lagrange multiplier, $\lambda$, and $\bar{\rho}^{2}$. A physical solution for all these parameters would be the one that minimizes $e_{g}$. 
We re-parameterize our mean-field parameters using the constraint in Eq.~\ref{cons2}.
\begin{subequations}
\begin{eqnarray}
 \left(\phi_{F},\alpha_{F}\right) &=& \sqrt{2}\,S\left(\cos\theta_{F},\sin\theta_{F}\right)\\
  \left(\phi_{A},\alpha_{A}\right) &=& \sqrt{2}\,S\left(\cos\theta_{A},\sin\theta_{A}\right)\\
   \left(\phi^\prime_{A},\alpha^\prime_{A}\right) &=& \sqrt{2}\,S\left(\cos\theta^\prime_{A},\sin\theta^\prime_{A}\right)
\end{eqnarray}
\end{subequations}

To solve for the physical values of the mean-field parameters, we minimise the following function using simplex method (see Ref.~\cite{Nelder65}).
\begin{align}\label{min_fn}
 & \mathcal F(\lambda,\bar{\rho}^{2},\theta_{F},\theta_{A},\theta^\prime_{A}) = e_{g} + w_{\lambda}\left(\frac{\partial{e_g}}{\partial{\lambda}}\right)^{2}+w_{\bar{\rho}^{2}}\left(\frac{\partial{e_g}}{\partial{\bar{\rho}^{2}}}\right)^{2}\nonumber\\
 & + w_{\theta_{F}}\left(\frac{\partial{e_g}}{\partial{\theta_{F}}}\right)^{2}
 + w_{\theta_{A}}\left(\frac{\partial{e_g}}{\partial{\theta_{A}}}\right)^{2} + w_{\theta^{\prime}_{A}}\left(\frac{\partial{e_g}}{\partial{\theta^\prime_{A}}}\right)^{2}
\end{align}
The tolerance of the standard deviation of the simplex is set to be $10^{-12}$. We set the weight, $w_{\lambda}$, to have a higher value than the other $w$'s. This ensures a faster convergence. It is also possible to get complex $E_{i\k}$ in the parameter space. This situation is avoided by adding a huge penalty to the function values. If the system becomes gapless then the minimization algorithm will not converge, or will give physically inconsistent results. In such cases, it will be required to introduce a condensation order-parameter for the gapless mode. This does not concerns us currently, as we find them to be gapped. 

\subsection{Results from the SBMFT Calculations}\label{subsec:SBMFT} 
The SBMFT formulated above allows us to investigate with increasing $J_{F}$ the following aspects of the problem: 1) spontaneous dimerization, if any, of the AFM bonds, and 2) the formation of spin-$1$ moments on the FM bonds.

\begin{figure}
\includegraphics[width=0.9\columnwidth]{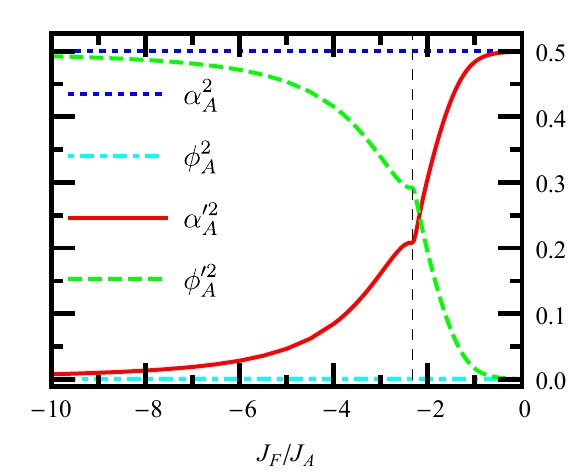}
\caption{\label{fig:hidaMF}
Variation of the mean-field parameters on the AFM bonds with $J_{F}$. The $\alpha_A$ and $\phi_A$ stay constant at $1/\sqrt{2}$ and $0$, respectively, for all $J_F$. The $\alpha^{\prime}_A$ and $\phi^{\prime}_A$ compete to dominate each other with a marked change in their relative strengths across $J_F/J_A \sim -2.33$. 
Notably, $\alpha^\prime_A \neq \alpha_A$ (except when $J_F/J_A$ is closer to zero), which implies spontaneous dimerization of the AFM bonds in Hida model.}
\end{figure}

Through the numerical minimization of the ground state energy, we compute the mean-field parameters as a function of $J_F$. On the FM bonds, we find $(\phi_F,\alpha_F)=(1/\sqrt{2},0)$ for all $J_F$. The other mean-field parameters are presented in Fig.~\ref{fig:hidaMF}. For the AFM bonds on the up-oriented hexagons, we find $(\phi_A,\alpha_A)=(0,1/\sqrt{2})$ for all $J_F$. The $\alpha^{\prime}_A$ and $\phi^{\prime}_A$ on the AFM bonds of the down-oriented hexagons, however, change with $J_F$ in an interesting way. For large negative $J_F$, the $\alpha^\prime_A$ is nearly zero, as opposed to $\alpha_A$. It clearly points to the dimerization of AFM bonds in Hida model. For very small values of $J_F$, $(\phi^\prime_A,\alpha^\prime_A)$ tends to become $(\phi_A,\alpha_A)$ [keeping ${\phi^{\prime}_A}^{2}+{\alpha^{\prime}_A}^{2}=1/2$, as per Eq.~\eqref{cons2}]. 
This suggests that for very small $J_F$, the dimerization tends to vanish and gives way to the uniform HS phase. In going from strong to weak negative $J_F$, the mean-field parameters vary continuously, but the slopes $d\phi^{\prime}_A/dJ_F$ and $d\alpha^{\prime}_A/dJ_F$ show a jump discontinuity at $J_{F}/J_{A}=-2.33$, which is an indication of a continuous quantum phase transition. 

For a more direct physical understanding of the SBMFT results, we calculate 1) the order parameter for the dimerization of the AFM bonds, and 2) the total spin moment per FM bond. The dimerization order-parameter, $O_D$, is defined as: 
\begin{equation}
O_{D}=\frac{1}{3N_{uc}}\left|\sum_{\langle i,j\rangle}^{\afup}\langle\vec{S}_{i}\cdot\vec{S}_{j}\rangle-\sum_{\langle i,j\rangle}^{\afdn}\langle\vec{S}_{i}\cdot\vec{S}_{j}\rangle\right|. 
\label{eq:OD}
\end{equation}
It distinguishes the singlet weight on the AFM bonds of the up-oriented hexagons from that of the down-oriented hexagons. The average total spin moment, $\widetilde{S}$, per FM bond is defined as follows:  
\begin{align}
 \widetilde{S}\left(\widetilde{S}+1\right)&=\frac{1}{3N_{uc}}\sum_{\langle i,j\rangle}^{\fer}\left\langle\left(\vec{S}_{i}+\vec{S}_{j}\right)^{2}\right\rangle\nonumber\\
 &=\frac{1}{3N_{uc}}\sum_{\langle i,j\rangle}^{\fer}\left(\langle\vec{S}_{i}^2\rangle+\langle\vec{S}_{j}^{2}\rangle+2\langle\vec{S}_{i}\cdot\vec{S}_{j}\rangle\right).
 \label{eq:moment}
\end{align}
To compute $O_D$ and $\widetilde{S}$, we rewrite Eqs.~\eqref{eq:OD} and~\eqref{eq:moment} in the Schwinger boson representation, and then calculate the expectation values in the ground state of $\mathcal{H}_{MF}^{SB}$.

\begin{figure}
\includegraphics[width=0.9\columnwidth]{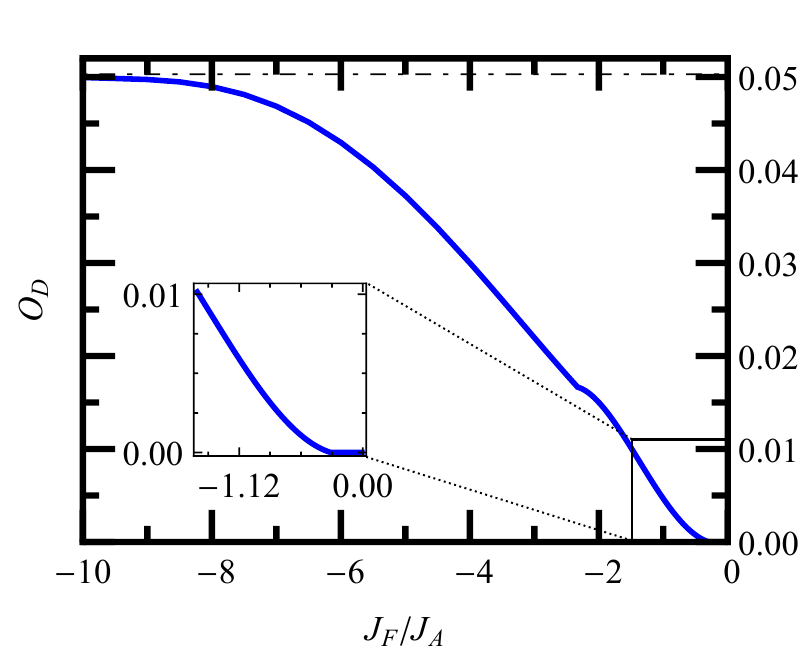}
\caption{\label{hidaTri}
The dimerization order parameter, $O_D$, vs. $J_F/J_A$. The horizontal (dot-dashed) line at $0.05$ corresponds to the value of trimerization order parameter for spin-$1$ kagom\'e antiferromagnet (see Appendix). The inset zooms in to show the dimerization transition at $J_F/J_A = -0.28$.} 
\end{figure}

The $O_D$ is found to be zero for very small values of $J_F$ implying a uniform HS phase. For $-J_F/J_A \gtrsim 0.28$, however, the $O_D $ takes non-zero values that grow continuously as shown in Fig.~\ref{hidaTri}. This implies ``spontaneous'' dimerization for the AFM bonds in Hida model. Thus, at $J_F/J_A = -0.28$, the SBMFT ground state of Hida model undergoes a symmetry-breaking transition from the HS to D-HS phase. Moreover, for large negative $J_F$, the $O_D$ is found to saturate to a value which is same as the ``trimerization" order-parameter value of the spin-1 KHA (see Appendix for the SBMFT calculation of the spin-1 KHA model). These findings are in qualitative agreement with what we learnt from TMFT.

\begin{figure}
\includegraphics[width=0.9\columnwidth]{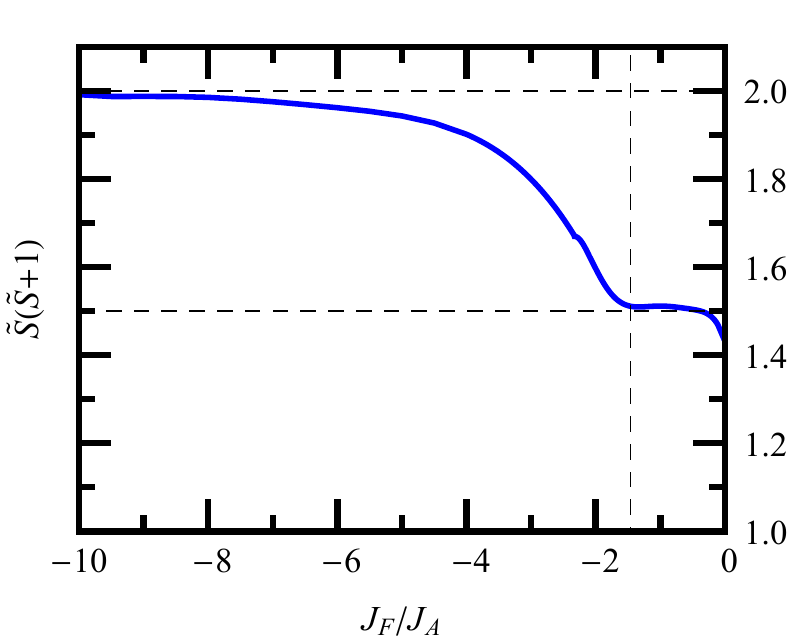}
\caption{\label{momJF}
The average total spin per FM bond vs. $J_F/J_A$. For large negative $J_F$, it approaches the value of a spin-1 moment. However, for $-J_F/J_A < 1.46$, it is (nearly) a constant at 1.5, which corresponds to having two uncorrelated spin-1/2's. It implies that only when $J_F/J_A$ is stronger than a critical value (-1.46) that the pair of spin-1/2's on a FM bond start to behave as a bound moment.} 
\end{figure}

The moment per FM bond, $\widetilde{S}$, correctly tends to 1 for large negative $J_F$, as shown in Fig.~\ref{momJF}. It decreases continuously as $|J_F|/J_A$ decreases. But for $|J_F|/J_A \lesssim 1.46$, the $\widetilde{S}(\widetilde{S}+1)$ stops decreasing and stays put at a value of 1.5, which corresponds to having two uncorrelated spin-1/2's on every FM bond~\footnote{Ideally, $\tilde{S}(\tilde{S}+1)$ should be exactly 1.5 at $J_F=0$. But in Fig.~\ref{momJF}, it goes a little below 1.5 for $J_F$ very close to zero. This minor discrepancy, we believe, can be cured by including the quantum fluctuations of the condensed modes, $\tilde{a}_-$ and $\tilde{b}_-$ of Eq.~\eqref{eq:FMbond}. In the present ``minimal'' formulation, we have ignored these fluctuations (which is  fine for $J_F$'s not so close to zero). }. 
This is an interesting result. It says that the spin-1/2's on the FM bonds of Hida model require a critical strength of $J_F$ to form bound moments! Thus, when $\widetilde{S}(\widetilde{S}+1)$ starts to increase from 1.5, it marks a transition from a phase of spin-1/2 moments to a phase with bound spin-1/2's on FM bonds. 

Therefore, according to our SBMFT calculations, two different quantum phase transitions occur in the ground state of Hida model. The first of these is the ``dimerization'' transition at $(J_F/J_A)_{c1} = -0.28$, across which the AFM bonds of Hida model undergo spontaneous dimerization. The second one is the ``moment-formation'' transition at $(J_F/J_A)_{c2} = -1.46$, under which the pair of spin-1/2's on every FM bond start expressing as a bound moment (that eventually becomes spin-1 for large $J_F$'s). This leads to the following three distinct phases: 
\begin{enumerate}
\item the uniform HS phase of spin-1/2 moments below $(J_F/J_A)_{c1}$, 
\item the D-HS phase of spin-1/2 moments between $(J_F/J_A)_{c1}$ and $(J_F/J_A)_{c2}$, and
\item the D-HS phase with bound moments on FM bonds above $(J_F/J_A)_{c2}$, which adiabatically continues to become the TS phase of spin-1 KHA for large negative $J_F$.
\end{enumerate}
In addition to this, as we noticed in Fig.~\ref{fig:hidaMF}, the jump discontinuities of the slopes of $\alpha^\prime_A$ and $\phi^\prime_A$ also suggest a quantum phase transition at $(J_F/J_A)_{c3}=-2.33$. But it is not clear how to interpret it, because $\alpha$'s and $\phi$'s are not physical observables. Besides, it has no particular bearing on the two physical transitions described above.  
But it seems to mark the qualitative change from a phase with weak dimerization of AFM bonds and weakly bound spin-1/2's on FM bonds to a  phase with strong dimerization and strongly bound moments. 

Overall, the SBMFT of Hida model presents a very novel picture of its ground state. It also adds nicely to the understanding of spontaneous trimerization in the ground state of the spin-1 KHA model.

\section{Summary}\label{sec:summary}
Motivated by recent studies on spin-$1$ kagom\'e Heisenberg antiferromagnet (KHA), we have investigated the Hida model, which is a spin-1/2 model of antiferromagnetic hexagons coupled via ferromagnetic bonds (on honeycomb lattice). We have employed triplon 
and Schwinger 
boson mean-field approaches to study the evolution of the ground state from the hexagonal singlet (HS) phase at small $J_F/J_A$  to the trimerized singlet (TS) phase at large negative $J_F/J_A$ (which is the ground state of spin-1 KHA).

From triplon mean-field theory we learnt that, at some intermediate value of $J_F/J_A$, the uniform HS ground state gives way to the dimerized hexagonal singlet (D-HS) state, which then remains the ground state of Hida model for all negative $J_F$'s. The TS ground state of spin-1 KHA is same as the D-HS ground state of Hida model at large negative $J_F$.

From the Schwinger boson mean-field theory, in an independent and unbiased way, we again found that the ground state of Hida model exhibits spontaneous dimerization at $J_F/J_A = -0.28$. It also revealed to us a second quantum phase transition at $J_F/J_A = -1.46$, under which the spin-1/2's of an FM bond begin to express as a bound moment, which gradually becomes spin-1 for stronger $J_F$. The dimerization order parameter in the ground state of Hida model approaches the same value as the trimerization order parameter for spin-1 KHA (see Appendix). Thus, both triplon and Schwinger boson methods produce a mutually consistent picture of the ground state of the Hida model, and tell us clearly about how trimerized singlet ground state is formed in a spin-1 KHA from the perspective of the Hida model. 

In the light of our investigations of the Hida model, we predict that the \ce{m-MPYNN.X} organic salts (which historically motivated these studies) would realize the D-HS phase at low temperatures, as opposed to the hexagonal singlet solid (HSS) phase considered by Hida. The D-HS phase is both non-magnetic and spin-gapped, which is qualitatively consistent with the known experimental features of these materials. But the same is also true of the uniform HS (or HSS)  phase. Therefore, we propose to ascertain the existence of dimerized hexagonal singlet phase in these organic salts by measuring the static structure factor using Neutron diffraction.

\begin{acknowledgments}
B.K. thanks Frederic Mila and Pierre Nataf for exciting discussions, and 
acknowledges the financial support under UPE-II scheme of JNU, DST-PUSRE and DST-FIST support for the HPC facility in SPS, JNU. P.G. acknowledges CSIR (India) for financial support. We also acknowledge IUAC (India) for using their HPC facility. 
\end{acknowledgments}

\appendix
\section{The SBMFT of Spin-$S$ kagom\'e Heisenberg antiferromagnet}
We present here a Schwinger boson mean-field study of the KHA. Although quite a few different SBMFT studies of KHA  have  been done before \cite{Sachdev1992, Manuel1994, Wang2006, Li2007, Messio2010, Mondal2017}, but they never explored the possibility of spontaneous trimerization therein. We try to fill this gap here. 
 
\begin{figure}
\includegraphics[width=0.9\columnwidth]{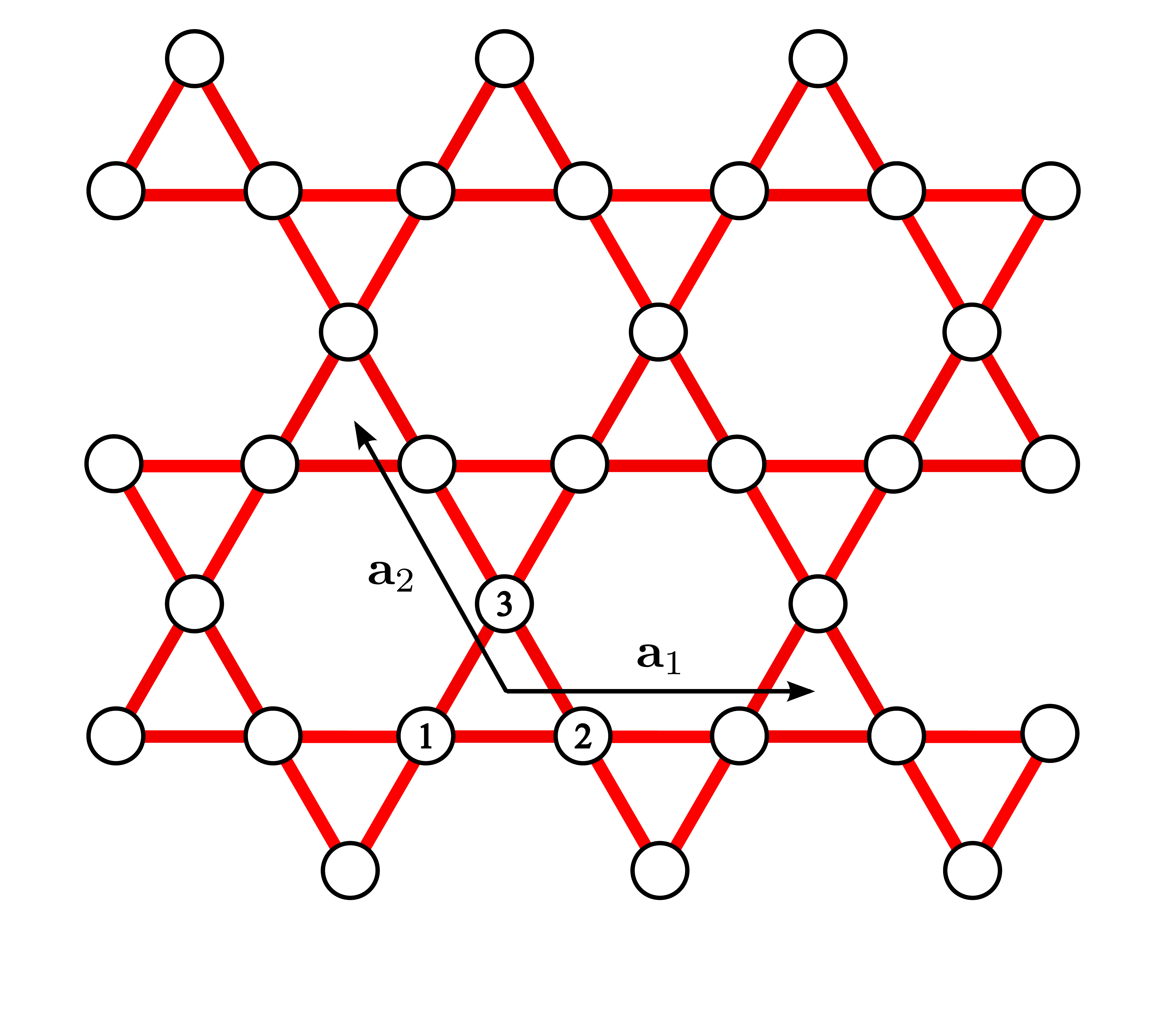}
\caption{\label{lat:Kag}
The kagom\'e lattice with primitive vectors $\a_{1}=2\hat{x}$ and  $\a_{2}=-\hat{x}+\sqrt{3}\hat{y}$. } 
\end{figure}

The spin-$S$ KHA model is given by
\begin{equation}
\mathcal{H}_{KHA}=\widetilde{J}_{A}\sum_{\langle i,j \rangle}{\bf S}_{i}\cdot{\bf S}_{j}, 
\end{equation}
where $\widetilde{J}_{A} > 0$ and ${\bf S}_{i}$'s are the spin-$S$ operators.  The spin-1 case with $\widetilde{J}_{A} = J_A/4$ is relevant to the Hida model for large $J_F$'s. In the Schwinger boson representation as defined in the main text, the KHA Hamiltonian reads as: 
\begin{eqnarray}
 \mathcal{H}_{KHA}=\frac{\widetilde{J}_{A}}{2}\left[\sum_{\langle i,j \rangle}^{\uptri}\left(:F^{\dagger}_{ij}F_{ij}:-A^{\dagger}_{ij}A_{ij}\right)\right.\nonumber\\
 +\left.\sum_{\langle i,j \rangle}^{\dntri}\left(:F^{\dagger}_{ij}F_{ij}:-A^{\dagger}_{ij}A_{ij}\right)\right].
\end{eqnarray}
In the mean-field approximation, with $\alpha=\langle A_{ij}\rangle$ and $\phi=\langle F_{ij}\rangle$  on the bonds of all \textit{up} triangles ($\uptri$, see Fig.~\ref{lat:Kag}), $\alpha^{\prime}=\langle A_{ij}\rangle$ and $\phi^{\prime}=\langle F_{ij}\rangle$ on the bonds on all \textit{down} triangles ($\dntri$), the Hamiltonian takes the following form:
\begin{eqnarray}
 \mathcal{H}^{SB}_{MF}=&&\frac{\widetilde{J}_{A}}{2}\sum_{\langle i,j \rangle}^{\uptri}\left[\phi\left(F^{\dagger}_{ij}+F_{ij}\right)-\alpha\left(A^{\dagger}_{ij}+A_{ij}\right)\right]\nonumber\\
 &+&\frac{\widetilde{J}_{A}}{2}\sum_{\langle i,j \rangle}^{\dntri}\left[\phi^{\prime}\left(F^{\dagger}_{ij}+F_{ij}\right)-\alpha^{\prime}\left(A^{\dagger}_{ij}+A_{ij}\right)\right]\nonumber\\
 &-&\frac{3\widetilde{J}_{A}}{2}N_{uc}\left[\left(\phi^{2}-\alpha^{2}\right)+\left(\phi^{\prime 2}-\alpha^{\prime 2}\right)\right]\nonumber\\
 &+&\lambda\left(\aidag a_{i}+\bidag b_{i}-2S\right).
\end{eqnarray}
Here, $\lambda$ is the Lagrange multiplier to satisfy the boson number constraint, $\aidag a_{i}+\bidag b_{i}=2S$, on average.

In momentum space, the mean-field Hamiltonian can be written as:
\begin{equation}\label{kagomemf}
\mathcal{H}_{MF}^{SB}=\sum_{\k}\left[ 
 \begin{array}{cc} {\bf{a}}_{\k}^{\dagger} & {\bf{b}}_{\k}\end{array}  
\right]\left[ \begin{array}{cc}
\mathcal{V}_{1\k} & \mathcal{V}_{2\k}   \\
\mathcal{V}_{2\k}^{\dagger} & \mathcal{V}_{1\k}  \end{array}\right]\left[ 
\begin{array}{c} {\bf{a}}_{\k}\\ {\bf{b}}_{\k}^{\dagger}  \end{array}  
\right]+e_{0}N_{uc}, 
\end{equation}
where, $ {\bf{a}}_{\k}^{\dagger} =\left[a_{1,\k}^{\dagger}\ a_{2,\k}^{\dagger}\ a_{3,\k}^{\dagger}\right]$ and $ {\bf{b}}_{\k}  =\left[b_{1,\k}\ b_{2,\k}\ b_{3,\k}\right]$. The Fourier transform of the Schwinger boson operators is defined below.
 \begin{eqnarray}
  a_{i}\left(\r\right)&=&\frac{1}{\sqrt{N_{uc}}}\sum_{\k}a_{i,\k} e^{-i\k.\r}\\
  b_{i}\left(\r\right)&=&\frac{1}{\sqrt{N_{uc}}}\sum_{\k}b_{i,\k} e^{i\k.\r}
 \end{eqnarray}
 In Eq.~\ref{kagomemf}
 \begin{equation}
  e_{0} = \frac{3\widetilde{J}_{A}}{2} \left[\left(\alpha^{2}-\phi^{2}\right)+\left(\alpha^{\prime 2}-\phi^{\prime 2}\right)\right] -  3\lambda\left(2S+1\right).
 \end{equation}
Moreover,
\begin{eqnarray}
\mathcal{V}_{1\k} &=& \frac{\widetilde{J}_{A}}{2\sqrt{2}}\left[ \begin{array}{ccc}
0 & \phi+\phi^{\prime}e^{i\k.\a_{1}} & \phi+\phi^{\prime}e^{i\k.\a_{3}}   \\
\phi+\phi^{\prime}e^{-i\k.\a_{1}} &  0 & \phi+\phi^{\prime}e^{i\k.\a_{2}}\\
\phi+\phi^{\prime}e^{-i\k.\a_{3}}& \phi+\phi^{\prime}e^{-i\k.\a_{2}} & 0 \end{array}\right]\nonumber\\
&&+\lambda\times\mathbb{I}_{3\times3}
\end{eqnarray}
and
\begin{equation}
\mathcal{V}_{2\k}=\frac{\widetilde{J}_{A}}{2\sqrt{2}}\left[ \begin{array}{ccc}
0 & -\alpha-\alpha^{\prime}e^{i\k.\a_{1}} & \alpha+\alpha^{\prime}e^{i\k.\a_{3}}   \\
\alpha+\alpha^{\prime}e^{-i\k.\a_{1}} &  0 & -\alpha-\alpha^{\prime}e^{i\k.\a_{2}}\\
-\alpha-\alpha^{\prime}e^{-i\k.\a_{3}}& \alpha+\alpha^{\prime}e^{-i\k.\a_{2}} & 0 \end{array}\right]
\end{equation}
with $\a_{3}=\a_{1}+\a_{2}$. 

The Hamiltonian in Eq. \ref{kagomemf} is diagonalized using Bogoliubov transformation. The ground state energy per unit cell in terms of the Bogoliubov quasiparticle dispersions, $E_{i,\k}$'s, is given by
\begin{equation}
 e_{g}^{KHA}=e_{0}+\frac{1}{2N_{uc}}\sum_{i,\k}E_{i,\k}.
\end{equation}
Given that $:F^{\dagger}_{ij}F_{ij}:+A^{\dagger}_{ij}A_{ij}=2S^{2}$, we reparameterize the mean-field variables as: $\left\{\alpha,\phi\right\} = \sqrt{2}\, S\left(\sin\theta,\cos\theta\right)$ and $\left\{\alpha^{\prime},\phi^{\prime}\right\} = \sqrt{2}\, S\left(\sin{\theta^\prime},\cos{\theta^\prime}\right)$.
We then numerically minimize the following weighted energy function with respect to $\lambda$, $\theta$ and $\theta^{\prime}$:
\begin{eqnarray}
 \mathcal F(\lambda,\theta,\theta^{\prime})=e_{g}^{KHA}&+&w_{\lambda}\left(\frac{\partial{e_g}}{\partial{\lambda}}\right)^{2}+w_{\theta}\left(\frac{\partial{e_g}}{\partial{\theta}}\right)^{2}\nonumber\\
  &+&w_{\theta^{\prime}}\left(\frac{\partial{e_g}}{\partial{\theta^{\prime}}}\right)^{2}.
\end{eqnarray}
To see spontaneous trimerization, if any, in the ground state, we define the trimerization order-parameter, $O_T$.

\begin{equation}
 O_{T}=\frac{1}{3N_{uc}}\left(\sum_{\langle i,j\rangle}^{\uptri}\langle\bf{S}_{i}\cdot\bf{S}_{j}\rangle-\sum_{\langle i,j\rangle}^{\dntri}\langle\bf{S}_{i}\cdot\bf{S}_{j}\rangle\right)
\end{equation}

\begin{figure}[b]
\includegraphics[width=0.9\columnwidth]{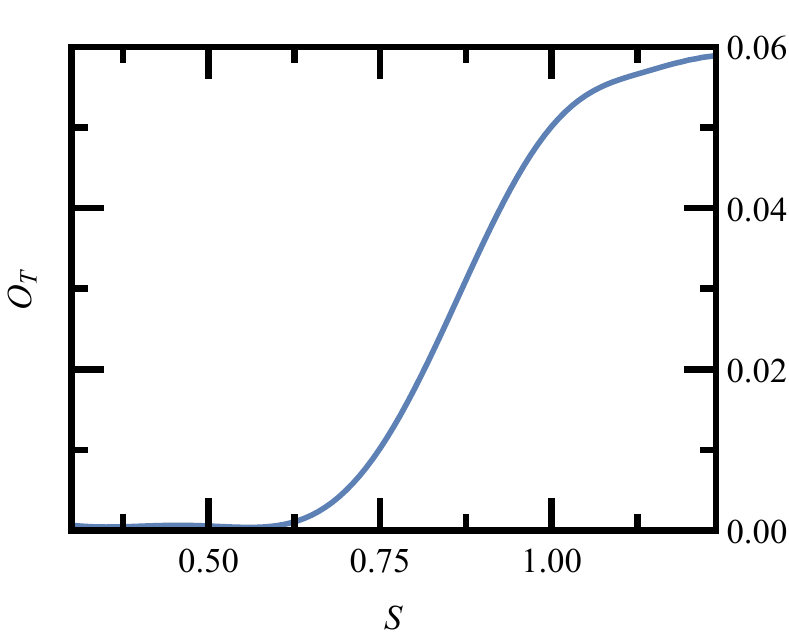}
\caption{\label{Kag_trimer}
 The trimerization order parameter, $O_T$, vs. $S$ from a Schwinger boson mean-field calculation of the spin-S KHA}. 
\end{figure}

The $O_{T}$ versus $S$, as obtained from this calculation, is shown in Fig.~\ref{Kag_trimer}. The data presented here corresponds to the ground state with gapped spin excitations. Here, we first see no trimerization ($O_T=0$) for $S \lesssim 0.62$. But then, for $S\gtrsim 0.62$, we obtain $O_T \neq 0$, which means spontaneous trimerization. It clearly implies the TS ground state from spin-1 KHA. Interestingly, our simple calculation also finds an expectedly different (uniform; without trimerization) ground state for spin-$1/2$ KHA. Moreover, for $S\gtrsim 1.24$, the SBMFT ground state is found to become gapless implying $\sqrt{3}\times\sqrt{3}$ AFM order. This is also consistent with the expected large $S$ behavior~\footnote{As the number fluctuations in SBMFT can give slightly lower value than $S(S+1)$ for the local moment, $\left < \vec{S}_i^2 \right >$~\cite{Messio2010},
it is possible that spin-3/2 KHA could also be gapped.}.

We end this discussion by noting that the $O_T=0.05$ for $S=1$ in Fig.~\ref{Kag_trimer} is same as the values of $O_D$ in Fig.~\ref{hidaTri} for large negative $J_F$. This shows that the trimerized singlet ground state of spin-1 KHA is adiabatically connected to the dimerized hexagonal singlet ground state of the Hida model.

\bibliography{references.bib}

\end{document}